\pdfoutput=1

\documentclass[a4paper,11pt]{article}

\usepackage{jheppub,fancyhdr,epstopdf,color,amsmath,cases,slashed,hyperref,subfigure}
\usepackage{amsfonts}
\usepackage{amssymb}
\usepackage{graphicx,graphics,epsfig}
\usepackage{geometry}
\usepackage[english]{babel}
\usepackage{mathtools}
\usepackage[utf8]{inputenc}
\usepackage[T1]{fontenc}
 \usepackage{float}
\usepackage{hyperref}
\usepackage{caption}
\usepackage{caption}
\usepackage{subfigure}  
\usepackage{mathrsfs}  
\usepackage{comment}
\usepackage{float}
\usepackage{stackrel}
\usepackage{multirow}
\usepackage{slashed}

\def\noi{\noindent}

\def\l{\lambda}

\def\ba{\begin{array}}
\def\ea{\end{array}}
\def\bea{\begin{eqnarray}}
\def\eea{\end{eqnarray}}

\newcommand{\Asla}{\not{\hbox{\kern-3.5pt $A$}}}
\newcommand{\Gsla}{\not{\hbox{\kern-3.5pt $G$}}}
\newcommand{\Wsla}{\not{\hbox{\kern-3.5pt $W$}}}
\newcommand{\Zsla}{\not{\hbox{\kern-3.5pt $Z$}}}

\newcommand{\Dslash}{\not{\hbox{\kern-4pt $D$}}}
\newcommand{\pslash}{\not{\hbox{\kern-2.3pt $p$}}}

\def\lsim{\;\raise0.3ex\hbox{$<$\kern-0.75em\raise-1.1ex\hbox{$\sim$}}\;}
\def\gsim{\;\raise0.3ex\hbox{$>$\kern-0.75em\raise-1.1ex\hbox{$\sim$}}\;}

\def\l{\lambda}

\def\ba{\begin{array}}
\def\ea{\end{array}}
\def\bea{\begin{eqnarray}}
\def\eea{\end{eqnarray}}

\def\bll{\tilde{\beta}_{\l_L}}

\def\lsim{\;\raise0.3ex\hbox{$<$\kern-0.75em\raise-1.1ex\hbox{$\sim$}}\;}
\def\gsim{\;\raise0.3ex\hbox{$>$\kern-0.75em\raise-1.1ex\hbox{$\sim$}}\;}

\newcommand{\beqn}{\begin{eqnarray}}
\newcommand{\eeqn}{\end{eqnarray}}

\title{Relic density of dark matter in the inert doublet model beyond leading order for the low mass region:  3. Annihilation in 3-body final state }
\preprint{LAPTH-003/21, CERN-TH-2021-003}

\author[a]{Shankha Banerjee}
\author[b]{\!\!, Fawzi Boudjema}
\author[c, d]{\!\!, Nabarun Chakrabarty}
\author[e]{\!\!,  Hao Sun}

\affiliation[a]{CERN, Theoretical Physics Department, CH-1211 Geneva 23, Switzerland}
\affiliation[b]{LAPTh, Universit\'e Savoie Mont Blanc, CNRS, BP~110, F-74941 Annecy-le-Vieux, France}
\affiliation[c]{Centre for High Energy Physics, Indian Institute of Science, C.V. Raman Avenue, Bangalore 560012, India}
\affiliation[d]{Department of Physics, Indian Institute of Technology Kanpur, Kanpur, Uttar Pradesh 208016, India}
\affiliation[e]{Institute of Theoretical Physics, School of Physics, Dalian University of Technology, Dalian 116024, People’s Republic of China}

\emailAdd{shankha.banerjee@cern.ch}
\emailAdd{boudjema@lapth.cnrs.fr}
\emailAdd{chakrabartynabarun@gmail.com}
\emailAdd{haosun@dlut.edu.cn}

\abstract{We perform the first one-loop electroweak corrections for  $2 \to 3$ processes for dark matter annihilation. These are the dominant processes that enter the computation of the relic density for the low mass region of the inert doublet model (IDM) when annihilations to two on-shell vector bosons are closed. The impact of the one-loop corrections are important as they involve, through rescattering effects, not only a dependence on the parameter controlling the dark sector, not present if a calculation at tree-level is conducted, but also on the renormalisation scale. These combined effects should be taken into account in analyses based on tree-level cross-sections of the relic density calculations, as a theoretical uncertainty which we find to be much larger than the cursory $\pm 10\%$ uncertainty that is routinely assumed, independently of the model parameters.}

\begin{document}

\date\today

\maketitle


\section{Introduction}
\label{sec:XXVff}
The fermions of the the standard model (SM) do not directly couple to the scalars in the inert doublet model (IDM)~\cite{Deshpande:1977rw, Barbieri:2006dq,Hambye:2007vf, LopezHonorez:2006gr,Cao:2007rm,Gustafsson:2007pc,Agrawal:2008xz,Hambye:2009pw, Lundstrom:2008ai, Andreas:2009hj, Arina:2009um, Dolle:2009ft, Nezri:2009jd, Miao:2010rg, Gong:2012ri, Gustafsson:2012aj, Swiezewska:2012eh, Arhrib:2012ia,Wang:2012zv, Goudelis:2013uca, Arhrib:2013ela, Krawczyk:2013jta, Osland:2013sla, Abe:2015rja,Blinov:2015qva, Diaz:2015pyv, Ilnicka:2015jba, Belanger:2015kga, Carmona:2015haa, Kanemura:2016sos,Queiroz:2015utg,Belyaev:2016lok,Arcadi:2019lka, Eiteneuer:2017hoh, Ilnicka:2018def, Kalinowski:2018ylg,Ferreira:2009jb,Ferreira:2015pfi,Kanemura:2002vm, Senaha:2018xek, Braathen:2019pxr, Arhrib:2015hoa,Garcia-Cely:2015khw,Banerjee:2016vrp,Basu:2020qoe,Abouabid:2020eik,Kalinowski:2020rmb}. The annihilation of the dark matter (DM) candidate in the IDM, the lightest neutral scalar $X$, occurs most naturally through annihilation into the SM vector bosons. These processes are triggered by the gauge coupling and also by the interactions stemming from the scalar sector of the model. The latter can be parametrised by the coupling of the SM Higgs to the pair of DM, $\l_L$, once the masses of all the scalars of the IDM are derived~\cite{OurPaper1_2020}. For these annihilations into a pair of $WW$ and $ZZ$ to be possible, the mass of the DM, $M_X$, must be larger than $M_W$, the mass of the $W$-boson. Even in this case, these annihilations are so efficient, see~\cite{OurPaper1_2020}, that the obtained relic density is too small, unless one considers very high DM masses~\cite{Banerjee:2019luv}. For the low mass, $M_X<M_W$, DM region of the IDM~\cite{OurPaper1_2020}, the annihilations are into $WW^\star$ and $ZZ^\star$ where one of the vector bosons is off-shell and is materialised by a fermion pair. The cross-sections are then smaller, bringing the relic density in accord with present measurement of the relic density. For $WW^\star$ and $ZZ^\star$, one is then faced with the calculation of a $2 \to 3 $ process which has never been attempted at one-loop for the calculation of the relic density.  Unlike the newly discovered co-annihilation region and the Higgs resonance region, this continuum region does not require much adjustment of the parameters in order to achieve a good value of the relic density within the freeze-out mechanism. This explains why a scan on the parameters of the IDM returns quite a few points with this topology for the relic density. Following the in-depth preliminary study, $XX \to W^+ W^-,ZZ$, of all the possible benchmarks in this region that pass all the experimental (and theoretical) constraints, we retain, in the present analysis of these channels, only those benchmarks which satisfy the one-loop perturbativity requirement~\cite{OurPaper1_2020}. This requirement was enunciated in the preparatory study $XX \to W^+ W^-,ZZ$~\cite{OurPaper1_2020}. In a nutshell, only models that return small enough $\bll$ (the $\beta$-function parameter that controls the running of the coupling $\l_L$) are perturbative~\cite{OurPaper1_2020}. On that basis, we keep three benchmark points defined in~\cite{OurPaper1_2020}: (points A, F and G) to illustrate our computations of the one-loop electroweak corrections to $XX \to Z f \bar f $ and $X X \to W f \bar f^\prime$.  Let us therefore recall the characteristics of these three benchmark points in Table~\ref{tab:bp2to3}.
\begin{table}[hbtp]
\centering
\begin{tabular}{|c|c |c |c|}
\cline{2-4}
\multicolumn{1}{c|}{}& A&F  & G\\
\hline
$M_X$  &   70&72& 72\\ 
\hline 
$\l_L \times 10^3$ &5.0 & 3.8&0.1  \\ \hline
$M_A$,$M_{H^\pm}$&      170,200&138,138& 158,158 \\  \hline 
$( \l_3,\l_4,\l_5)$ & (1.127, -0.738,-0.384) & (0.447,-0.222,-0.222)& (0.632,-0.316,-0.316)\\ 
\hline$\bll$ & $4.343+ 6.082 \l_2      $& $1.402 + 2.704 \l_2 $ & $2.142 + 3.795 \l_2 $ \\ \hline
\hline
   \multicolumn{4}{|c|}{$\Omega h^2$} \\    \hline
$\alpha(0)$ &  0.156&0.119 &0.121  \\    \hline
    $\Omega_{WW^\star}(\%)$      &90&88  &88 \\ \hline
     $\Omega_{Z Z^\star}(\%)$ & 10&12 &12 \\ 
 \hline
 \end{tabular}
\caption{\label{tab:bp2to3}{\it Characteristics of the benchmark points A,F and G. All masses are in GeV. The tree-level (calculated with $(\alpha(0)$) relic density and the weight in percent of each channel contributing to the relic density, are given. We also list the values of the underlying parameters $\l_{3,4,5}$ and $\bll$~\cite{OurPaper1_2020}.}}
\end{table}

The paper is organised as follows. In the next section, we review the $2 \to 3$ cross-sections and seek a factorisation where the flavour dependence is carried by the vector bosons' partial widths. Section~\ref{sec:oneloop2to3general} is a general presentation of the one-loop calculation. Since the bulk of the corrections are contained in the purely virtual correction in the neutral channel $Z \nu \bar \nu$, section~\ref{sec:znunuoneloopall} is dedicated to this channel before studying in section~\ref{sec:XXtoVffone-loop} all the other channels where final state radiation (tree-level $2 \to 4$ processes are needed) is considered. Section~\ref{sec:intermediatesummary} summarises all the one-loop results on the cross-sections leading the way to the impact of the corrections and the scale uncertainty on the relic density which we present in section~\ref{sec:reliconeloop}. Our conclusions are presented in section~\ref{sec:conclusions}.


\section{Tree-level considerations}
\label{sec:treelevel2to3}
\begin{figure}[hbtp]
\begin{center}
\includegraphics[width=0.9\textwidth, height=0.4\textwidth]{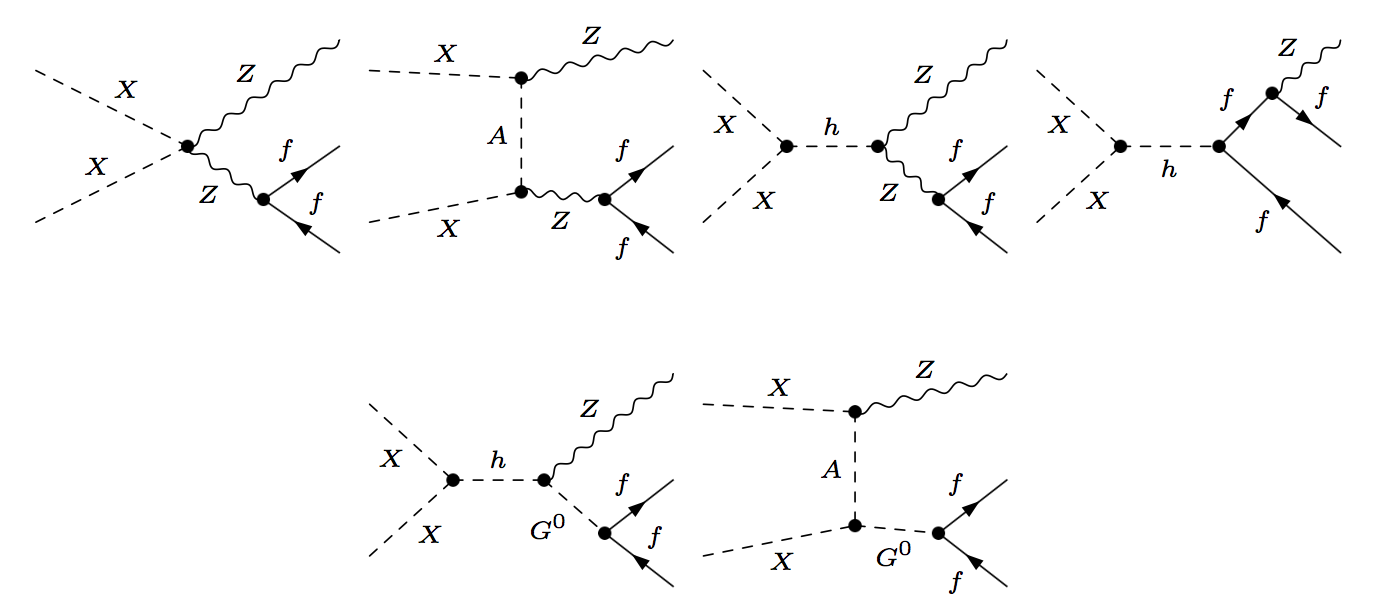}
\caption{\label{fig:fendiagXXVVtree}\it A selection of tree-level Feynman diagrams for $XX \to Z f \bar f$ in the Feynman gauge. The displayed diagrams can be built up from $XX \to Z Z^\star \to Z f \bar f$ but note also the $Z$-"bremsstrahlung" contribution triggered from $XX \to f f^\star \to f \bar f Z$ (last diagram in the first row).}
\end{center}
\end{figure}

Like for the $2 \to 2$ processes, $XX \to WW$ and $XX \to ZZ$, beside the masses of the dark sector particles, the cross-sections depend not only on the gauge coupling but also on $\l_L$ (because of the SM Higgs exchange and the quartic $XXVV$ couplings, $V=W,Z$). The massive fermions' Yukawa couplings may also play a role, but we will see that they are negligible. As a subset, contributions to the full $Z f \bar f$ is displayed in Figure~\ref{fig:fendiagXXVVtree}.\\
\noindent It is completely unwise to try to compute such cross-sections, even at tree-level, by splitting them into a $2 \to 2$ process followed by the ``decay" of one of the vector bosons into fermions even if the $Z/W$ current is conserved in the limit $m_f \to 0$. For starters, $XX \to V V^\star$ is ill-defined since it does not correspond to an element of the $S$-matrix. Therefore, a complete $2 \to 3, XX \to Z f \bar f$ and $ XX \to W f \bar f^\prime$, calculation is in order. Nonetheless, because the mass of the final fermions is very small compared to the energies involved and the fact that the fermions do not couple to the dark sector, we expect that the complete calculation of the cross-sections are arranged such that  
\beqn
\label{eq:approxZffwidth}
\frac{\sigma(XX \to Z f \bar f)}{\sigma(XX \to Z \nu \bar \nu)}  &\simeq &\frac{\Gamma_{Z \to f \bar f }}{\Gamma_{Z \to \nu \bar \nu }}=\frac{{\rm Br}_{Z \to f \bar f }}{{\rm Br}_{Z \to \nu \bar \nu }},  \\
\label{eq:approxWffwidth}  \frac{\sigma(XX \to W f \bar f^\prime)}{\sigma(XX \to W \nu_e \bar e)} 
& \simeq & \frac{\Gamma_{W \to f \bar f ^\prime}}{\Gamma_{W \to \nu_e \bar e }} \simeq N_c^f , \quad (N_c^f=1,3 \;\; \textrm{is the colour factor}).
\eeqn
\begin{center}
\begin{figure}[hbt]
\begin{center}
\includegraphics[width=0.48\textwidth,height=0.36\textwidth] {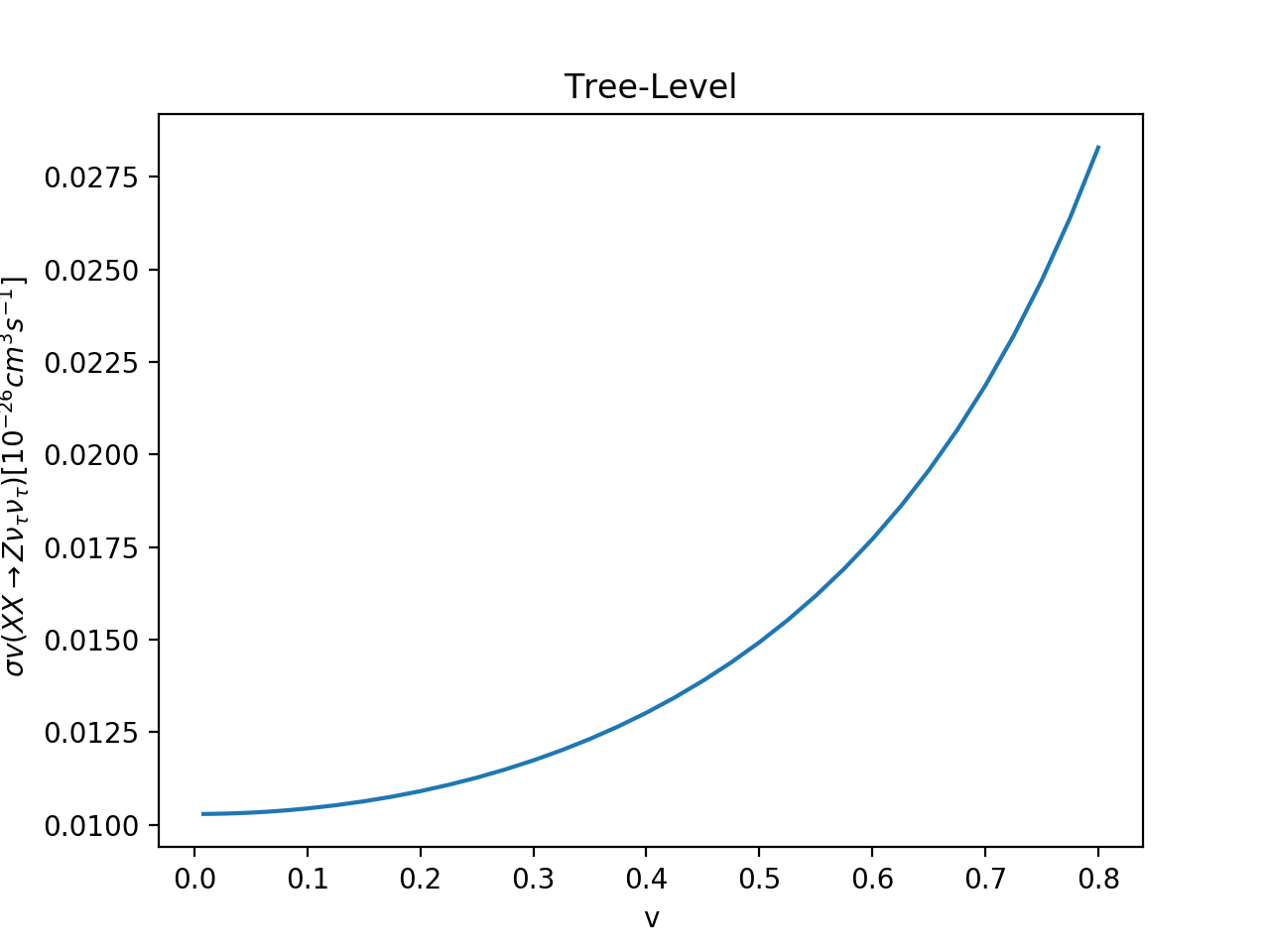}
\includegraphics[width=0.48\textwidth,height=0.36\textwidth]{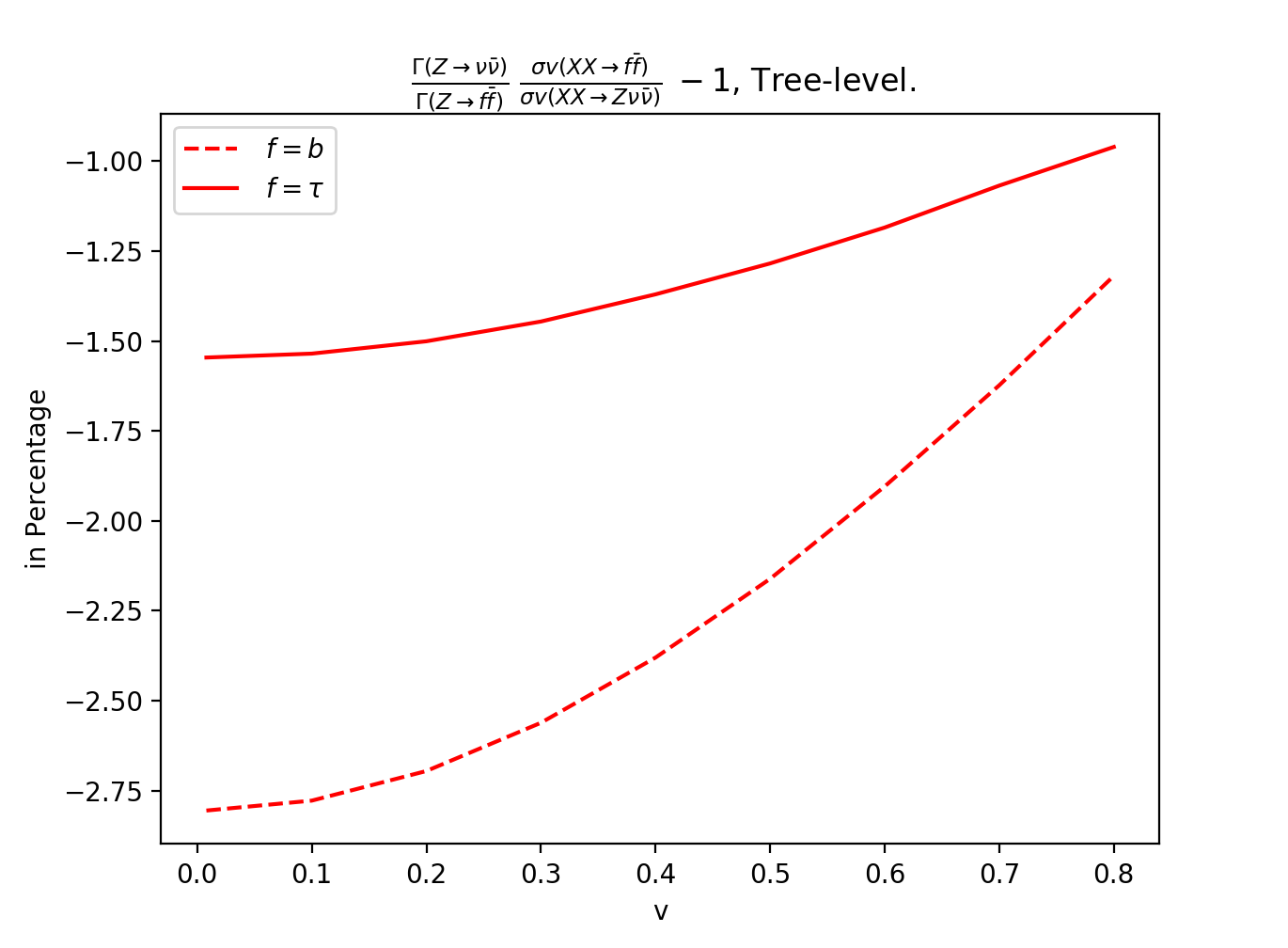}
\includegraphics[width=0.48\textwidth,height=0.36\textwidth]{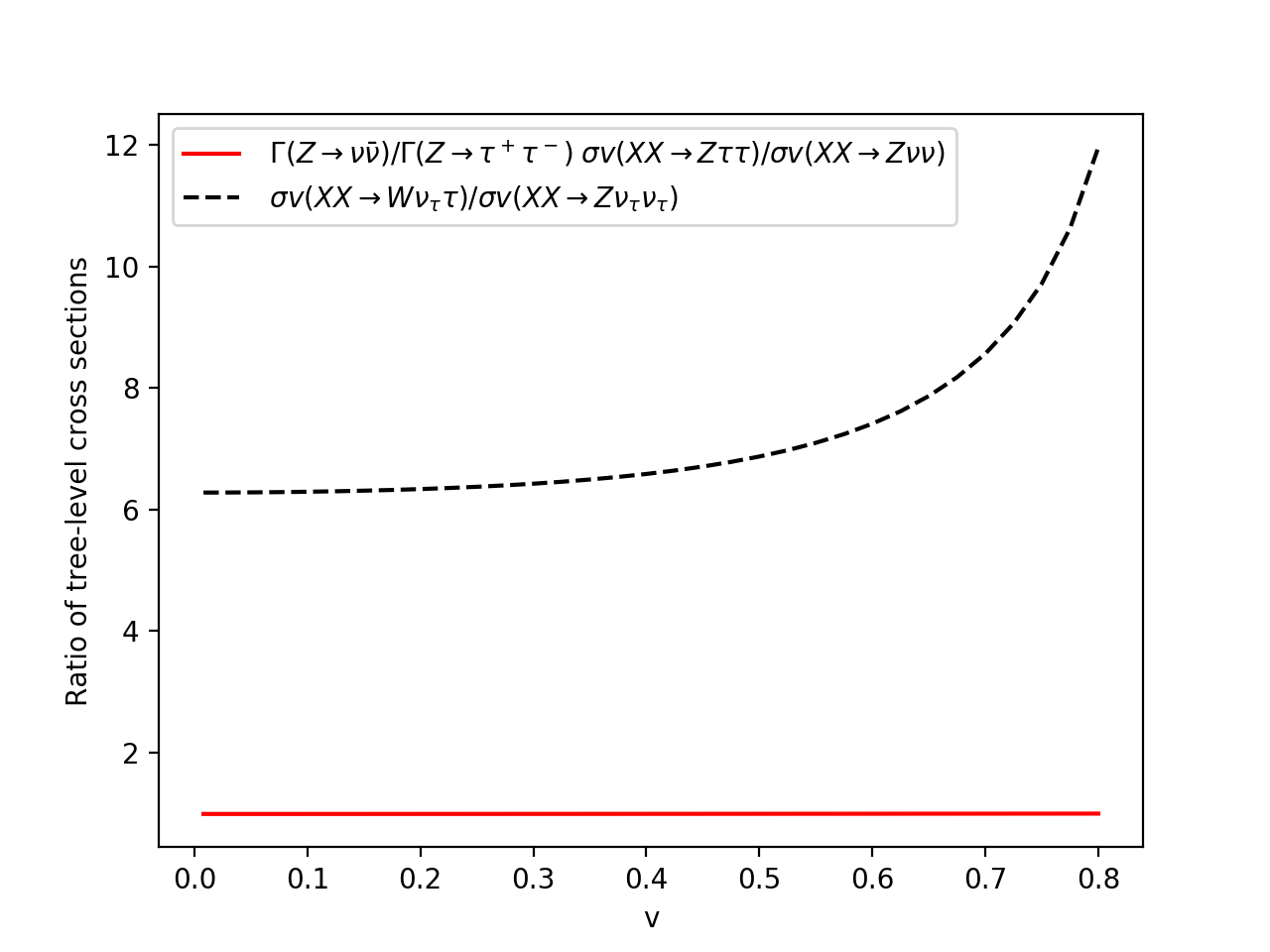}
\caption{\label{fig:relativeweightXXVVstar}\it Benchmark {\bf Point G}. The tree-level cross-section times the relative velocity for $XX \to Z \nu \bar \nu$ (upper-left panel), $R^{XX}_{Zf \bar f}-1$ for $f=\tau,b$ (upper-right panel). The lower panel displays $R^{XX}_{Z\tau \bar \tau} \simeq 1$ and the ratio of the tree-level cross-sections $R^{XX}_{W/Z}=\sigma (XX \to W \tau \bar{\nu}_\tau)/\sigma (XX \to Z \nu_\tau  \bar \nu_\tau)$ as a function of the relative velocity, $v$.}
\end{center}
\end{figure}
\end{center}
The ratios of the partial physical widths act as a normalisation of the cross-sections with respect to the neutrino channels. If we introduce 
\beqn
\label{eq:RatioapproxZffwidth}
R^{XX}_{Zf \bar f}& =&  \frac{\sigma(XX \to Z f \bar f)}{\sigma(XX \to Z \nu \bar \nu)} \; \frac{\Gamma_{Z \to \nu \bar \nu }}{\Gamma_{Z \to f \bar f }} \nonumber \\\label{eq:RatioapproxWffwidth} R^{XX}_{W f \bar f^\prime} &=&\frac{\sigma(XX \to W f \bar f^\prime)}{\sigma(XX \to W \nu_e \bar e)} \;  \frac{\Gamma_{W \to \nu_e \bar e }}{\Gamma_{W \to f \bar f ^\prime}}, \nonumber \\
R^{XX}_{W/Z}& =&\frac{\sigma(XX \to W \nu_e \bar e)}{\sigma(XX \to Z \nu \bar \nu)}, 
\eeqn
it follows from the arguments that lead to Equations~\ref{eq:approxZffwidth}-~\ref{eq:approxWffwidth} that $R^{XX}_{Zf \bar f} \sim R^{XX}_{W f \bar f^\prime} \sim 1$. We verify these approximations and behaviour by carrying a full calculation with fermion mass effect for the different channels. For Point G, the results are displayed in Figure~\ref{fig:relativeweightXXVVstar}.
 
First of all, the velocity dependence of the tree-level $XX \to Z f \bar f $ and $XX \to W f \bar f^\prime$ is strong. This is shown in Figure~\ref{fig:relativeweightXXVVstar} for the $Z \nu \bar \nu$ and consequently for $W \bar{\nu}_\tau \tau$. The latter grows faster past $v \sim 0.5$. This is understandable since as $v$ increases one gets closer to the opening of the threshold for on-shell pair production of vector bosons, the $WW$ threshold occurring first. The important observation though is that below $v\sim 0.5$ (the most important range for the relic density calculation), when the threshold effect is small, the ratio between these two cross-sections is almost constant. This is due to the global $SU(2)$ symmetry with the important consequence that the $\l_L$ dependence is the same in the neutral and charged channels. This is the same property that is explicitly confirmed in $XX \to WW$ and $XX \to ZZ$ past the $ZZ$ threshold in our study in Ref.~\cite{OurPaper1_2020}. The same $\l_L$ dependence will mean that both channels will exhibit the same scale uncertainty in the one-loop corrected cross-sections.\\
\noi We confirm this feature at all $v$, and independently of the flavour $R^{XX}_{W f \bar f^\prime}=1$, at better than the per-mille level, to the point where we can not display this difference in Figure~\ref{fig:relativeweightXXVVstar}.\\
\noi In the neutral channel, $R^{XX}_{Zf \bar f} \simeq 1$ is also nicely confirmed. Departure from unity are largest for the $Z b \bar b$ final state where the maximal value is  below $3\%$ across all values of $v$. The effect is smaller still for the $Z \tau \tau$ channel as seen in Figure~\ref{fig:relativeweightXXVVstar}. For all other channels, the mass effects are unnoticeable and are therefore not shown in the figure. The effect of the fermion masses/Yukawa couplings through Higgs exchange, $XX \xrightarrow{h} f f^\star \to f\bar{f}Z$, is therefore small.\\
\noi Taken together, these observations lead us to conclude that the cross-section into the neutrinos can be taken as a representative of the channels, neutral and charged. We also look at $R^{XX}_{W/Z}$. For the moment we keep these observations in mind before attempting the one-loop analyses. This first exploration confirms that the neutrino channels carry the bulk of the $v$ dependence and are the prime channels against which we will measure all other channels. We also confirm that the same conclusions, with the same level of accuracy, apply  for the other benchmark points (A and F). 


\section{$XX \to W f \bar f$ and $XX \to Z f \bar f$ at one-loop: General issues}
\label{sec:oneloop2to3general}
\begin{center}
\begin{figure}[bthp]
\begin{center}
\includegraphics[width=0.85\textwidth,height=0.55\textwidth] {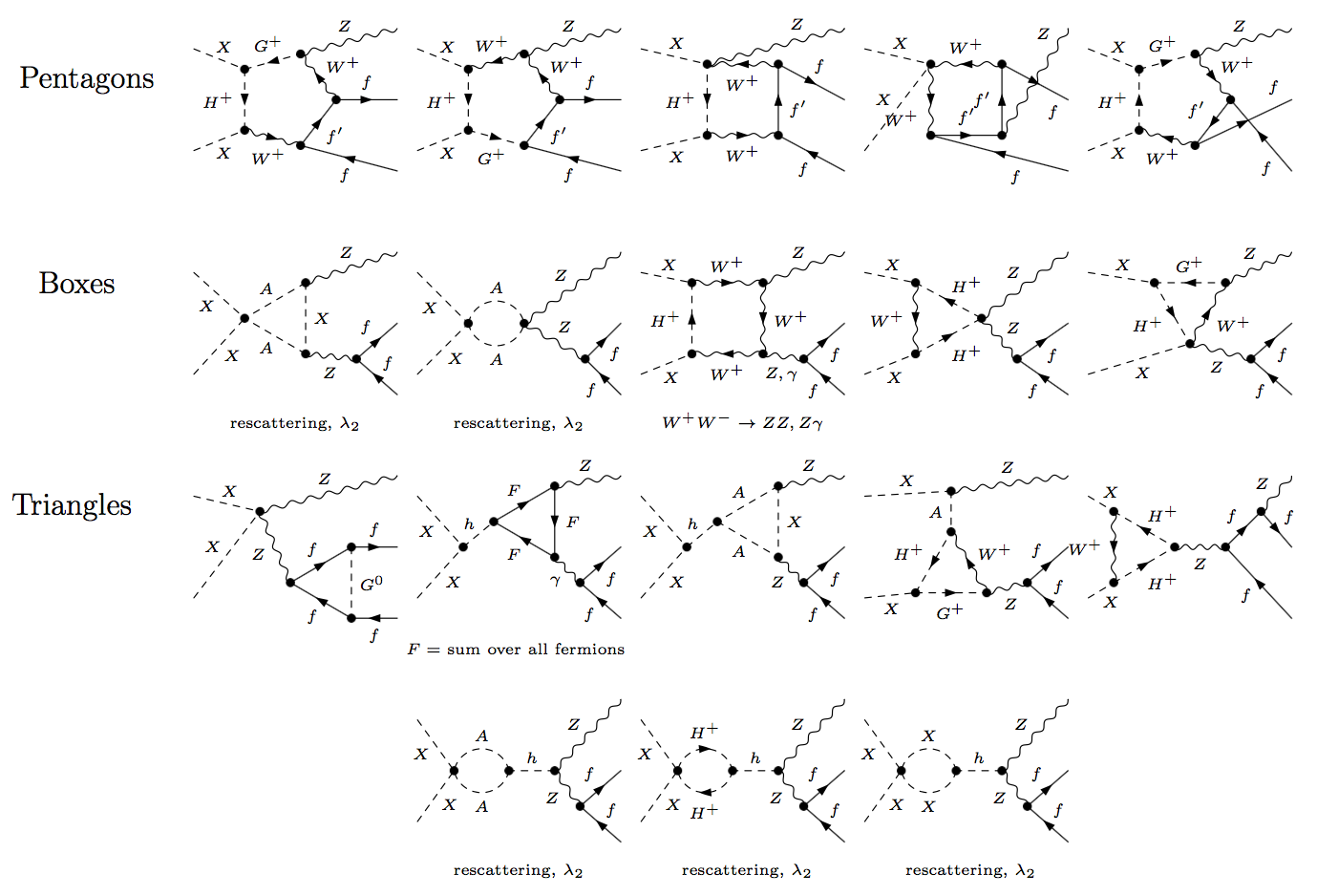}
\end{center}
\caption{\label{fig:fendiagXXVVLoop}\it A small selection of Feynman diagrams for $XX \to Z f \bar f $ at one-loop. We only show a very small subset of pentagons, boxes and triangles but not self-energy corrections and counterterms. $f ^\prime$ stands for the $SU(2)$ partner of $f$ ($\nu_l$ for $f=l$). $F$ stands for the sum of all SM fermions. Although some diagrams may not look like boxes, they fall under the box category because of the 4-particle vertices they involve. The same applies to triangles. We see rescattering effects within the dark sector, $XX \to AA, H^+ H^-, XX$ that explicitly involve the $\l_2$ parameter that does not show up at tree-level. Note also that because of the off-shell $Z$, charged fermions $ff$ pairs from $\gamma^\star$ must also be taken into account.}
\end{figure}
\end{center}

At one-loop, a large number of topologies appears for these $2 \to 3$ processes. For $XX \to Z f \bar f$, a set of the contributing one-loop diagrams is shown in Figure~\ref{fig:fendiagXXVVLoop}. A subset of diagrams at one-loop for $XX \to W f \bar f$ is found in our analysis of the co-annihilation region where this cross-section was a subdominant contribution to the relic density~\cite{OurPaper2_2020}. Technically, it is not (only) the sheer number of diagrams that adds to the complexity of the calculation but also the fact that the calculation, especially the reduction of the $n-$point integrals, is very much computer-time consuming. This is particularly the case for the $5-$point functions, the pentagons, that need to be evaluated, for instance, in configurations of phase space where $v\sim 0$, dangerously close to the appearance of very small Gram determinants, see~\cite{Boudjema_2005}. For the charged final state, tree-level $XX \to Z f \bar f \gamma$ and $XX \to W f \bar f \gamma$ must be considered together with the virtual one-loop corrections. We find that the phase space slicing method, as applied in~\cite{Banerjee:2019luv}, converges relatively quickly. Gauge parameter independence~\cite{OurPaper1_2020} is carried at some random point in phase space. This is a check not only on the model implementation (including counterterms) but also the tensorial reduction of the loop integrals.

An on-shell scheme for $\l_L$ based on $h \to XX$ is not possible for this mass range since this decay is closed. The radiative correction will therefore be sensitive to the renormalisation scale, $\mu$, associated with the $\overline{\text{MS}}$ scheme associated with a definition of $\l_L$, see~\cite{OurPaper1_2020}, 

\beqn
\label{deltalLMS}
\delta^{\overline{{\rm MS}}} \l_L =\frac{1}{32 \pi^2} \tilde{\beta}_{\l_L} C_{{\rm UV}}, \quad C_{{\rm UV}}=-\frac{2}{\varepsilon} - 1+\gamma_E-\ln(4\pi),
\eeqn
where $\varepsilon=4-d$ with $d$ being the number of dimensions in dimensional regularisation and $\gamma_E$ being the Euler-Mascheroni constant. With $\mu_{\text{dim}}$, the scale introduced by dimensional regularisation, the scale, $Q^2$, dependence of $\l_L$ is 
\beqn
\label{eq:delmulL}
32 \pi^2 \frac{\partial \l_L}{\partial \ln(Q^2)}=-32 \pi^2 \frac{\partial \l_L}{\partial \ln(\mu_{\text{dim}}^2)}=\tilde{\beta}_{\l_L}.
\eeqn

We will, for the three benchmark points in this mass range, study the scale dependence. Beside the scale dependence, we also investigate the interesting $\l_2$ dependence which, as we showed in $XX \to WW, ZZ$ above threshold, is not totally contained in $\beta_{\l_L}$, see~\cite{OurPaper1_2020}. The scale dependence, for a fixed $v$, is easily extracted from the $\l_L$ dependence of the tree-level cross-section combined with the expression of $\beta_{\l_L}$ which is known analytically. Barring very small mass effects, we confirm that at tree-level $R^{XX}_{Zf \bar f} \simeq 1,R^{XX}_{W f \bar f^\prime} =1$, while $R^{XX}_{W/Z}$ has a slight $v$ dependence above $v>0.5$. This is an indication that the $\l_L$ dependence of the cross-section is essentially the same for all fermion channels in $XX \to Z f \bar f$ and $XX \to W f \bar f^\prime$. 

Before displaying the numerical results of the full one-loop computation, we present the analytical scale variation for a chosen relative velocity in order to weigh how strong the scale dependence can be. From what we argued and will confirm shortly through a full calculation, the bulk of the scale variation is almost flavour independent. We therefore first concentrate on the neutral channel $XX \to Z \nu \bar \nu$. The most important features that are present in all other channels, are revealed in this channel. From the computation point of view, the neutrino channel is somehow the easiest since we do not need to deal with the infrared singularities that require the inclusion of the radiative $2\to 4 (3+\gamma)$ tree-level contribution. 


\section{$XX \to Z Z^\star$ at one-loop: $X X\to Z \nu \bar{\nu}$}
\label{sec:znunuoneloopall}
The large number of diagrams and the appearance of 5-point function loop integrals makes these computations challenging but {\tt SloopS}~\cite{Boudjema_2005, Baro:2007em,Baro:2008bg, Baro:2009na, Boudjema:2011ig, Boudjema:2014gza, Belanger:2016tqb, Belanger:2017rgu, Banerjee:2019luv}, our automated code, has been optimised to deal with many of the technicalities that are involved in these calculations. Another sort of technicality is the renormalisation and in particular the scheme dependence. All parameters but, in this case, $\l_L$ are defined on-shell. $\l_L$ is here taken $\overline{{\rm MS}}$, and at the end the one-loop result carries a scale dependence. As shown in details in the accompanying paper~\cite{OurPaper1_2020}, the scale dependence in this mixed scheme only originates from the $\l_L$ counterterm. One can even exactly determine the scale dependence of the {\underline{one-loop}} cross-section from the parametric $\l_L$ dependence of the {\underline{tree}}-level cross-section and the knowledge of the corresponding $\beta$ function for $\l_L$, $\bll$, which can be derived analytically. That such an approach agrees with the result of a direct calculation for an arduous calculation such as this $2\to 3$ process, is a further strong indication of the correctness of the calculation beside the tests of ultra-violet (UV) finiteness and gauge parameter independence. Moreover, such an approach which allows an analytical parametrisation of the scale, is very useful. The first step is to seek the $\l_L$ parametric dependence of the tree-level cross-section at a given $v$. To extract this, we maintain all parameters of the model, namely the masses and the SM parameters fixed apart from $\l_L$. Since the dependence is a quadratic polynomial, the $\l_L$ dependence is reconstructed numerically by generating the cross-sections for $\l_L=0,1,2$. We check the goodness of the parameterisation by taking a random value of $\l_L$ and comparing the cross-section obtained from the reconstructed polynomial against a direct calculation using the code. We always find excellent agreement for this check. One can then derive the infinitesimal change of the cross-section due to an infinitesimal change of $\l_L$. The latter will then be turned to a change due to the counterterm for $\l_L$ through $\bll$ which quantifies the scale dependence as we will see next for our three benchmark points. 

\subsection{$X X\to Z \nu \bar{\nu}$ at Point G} 
The $\l_L$ dependence of the cross-section, for $v=0.4$, is found to be  
\beqn
\sigma_{XX\to Z \nu \bar \nu, G}^0(v=0.4)\;v=0.013 + 0.300\l_L + 1.750\l_L^2 \simeq 0.013 (1+23 \l_L+135\l_L^2).
\eeqn 
Observe that the $\l_L$ dependence is quite strong. The (relative) coefficient of $\l_L$ is about $23$. Note also that the $\l_L^2$ is even larger. The latter will not be so important since the constraints posed on $\l_L$ give very small values of $\l_L$.  
One can then relate the one-loop correction ${\rm d}\sigma(\mu_2)$ at scale $\mu_2$ to the one at scale $\mu_1$ according to
\beqn
\label{eq:sigmazzstarmudepI}
{\rm d}\sigma_{XX\to Z \nu \bar \nu}^{G,v=0.4}(\mu_2)\;v&=& {\rm d}\sigma_{XX\to Z \nu \bar \nu}^{G, v=0.4}(\mu_1) v
- \bigg(0.300 + 2\times 1.750 \l_L \bigg)\frac{\beta_{\l_L} \ln(\mu_2/\mu_1)}{16\pi^2}, \nonumber \\
&\sim& {\rm d}\sigma_{XX\to Z \nu \bar \nu}^{I, v=0.4}(\mu_1)v\;-\;0.300 \; \bigg(2.141+3.795 \l_2 \bigg) \frac{\ln(\mu_2/\mu_1)}{16\pi^2}, \quad \text{leading to}\nonumber \\
\left. \frac{{\rm d}\sigma_{XX\to Z \nu \bar \nu}(2 \mu_1)}{\sigma^0_{XX\to Z \nu \bar \nu}} \right\rvert_{G,v=0.4}&\sim& \Bigg(\left. \frac{{\rm d}\sigma_{XX\to Z \nu \bar \nu}(\mu_1)}{\sigma^0_{XX\to Z \nu \bar \nu}} \right\rvert_{I, v=0.4}\;-\;21.7 \; (1+1.8\l_2) \Bigg)\%.
\eeqn
The last line gives the difference when the scale is doubled from $\mu_1$ to $2 \mu_1$. \\
\noi We verify these formulae against the results of a direct computation of the full one-loop correction. We obtain a 5 digit agreement for three values of $\l_2$ ($\l_2=0.01,1,2$) and different combinations of the scale $\mu$. We note that the scale dependence is quite large. This is due to the strong $\l_L$ dependence of the cross-section and also to the fact that $\bll$ is not so small. For this benchmark point and for $v=0.4$ we learn from Equation~\ref{eq:sigmazzstarmudepI} that, in the range $M_X/2<\mu< 2 M_X$, the uncertainty introduced by the scale variation is about $44\%$ for $\l_2=0$, it increases to about $200\%$ for $\l_2=2$, which should be quoted as the overall theoretical uncertainty if we allow both the scale $\mu$ to span the range $M_X/2-2M_X$ and a variation $0<\l_2<2$.
\begin{center}
\begin{figure}[bthp]
\begin{center}
\includegraphics[width=0.65\textwidth, height=0.4\textwidth]{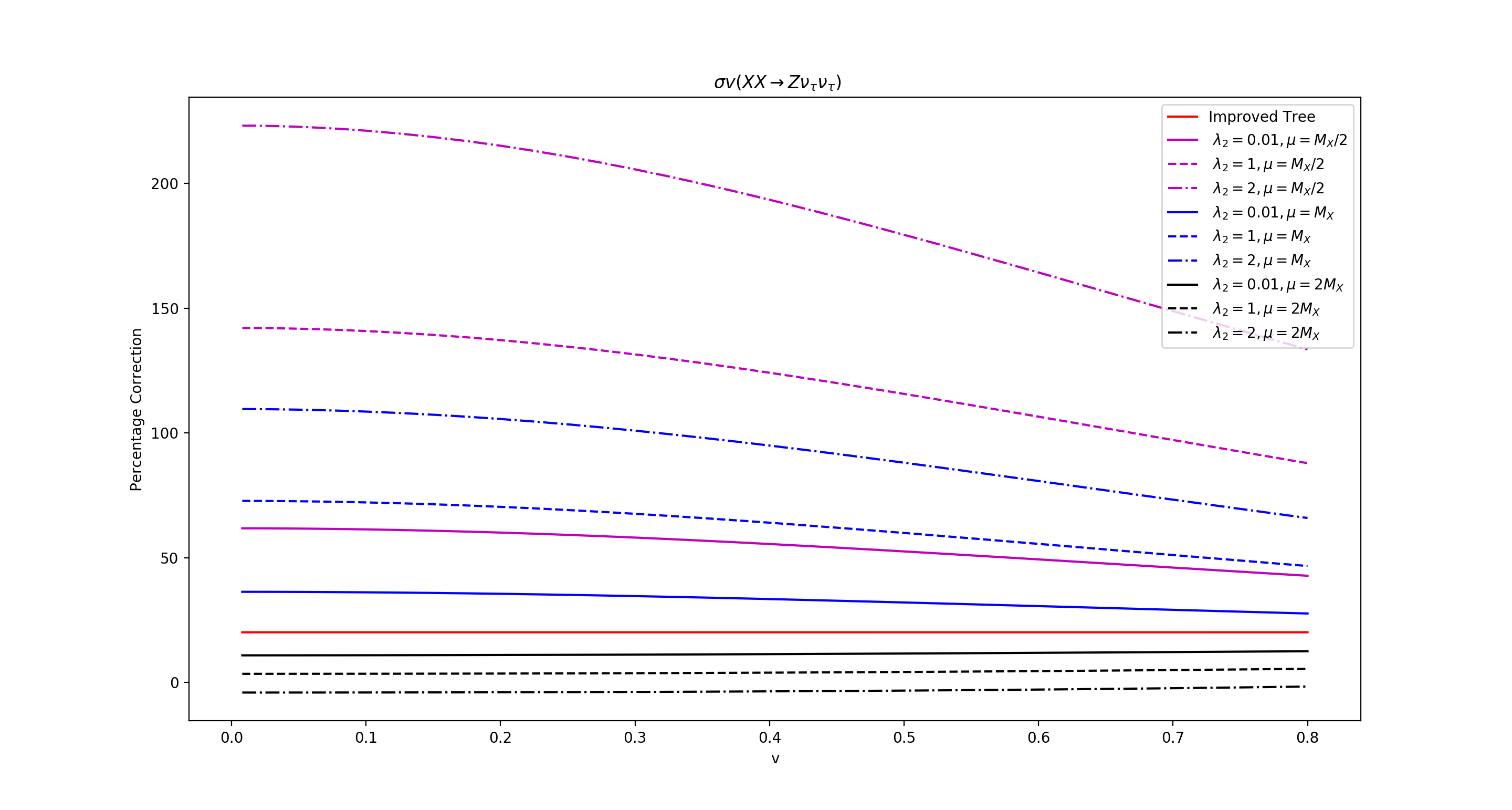} \hspace*{-0.5cm}
\includegraphics[width=0.34\textwidth, height=0.4\textwidth]{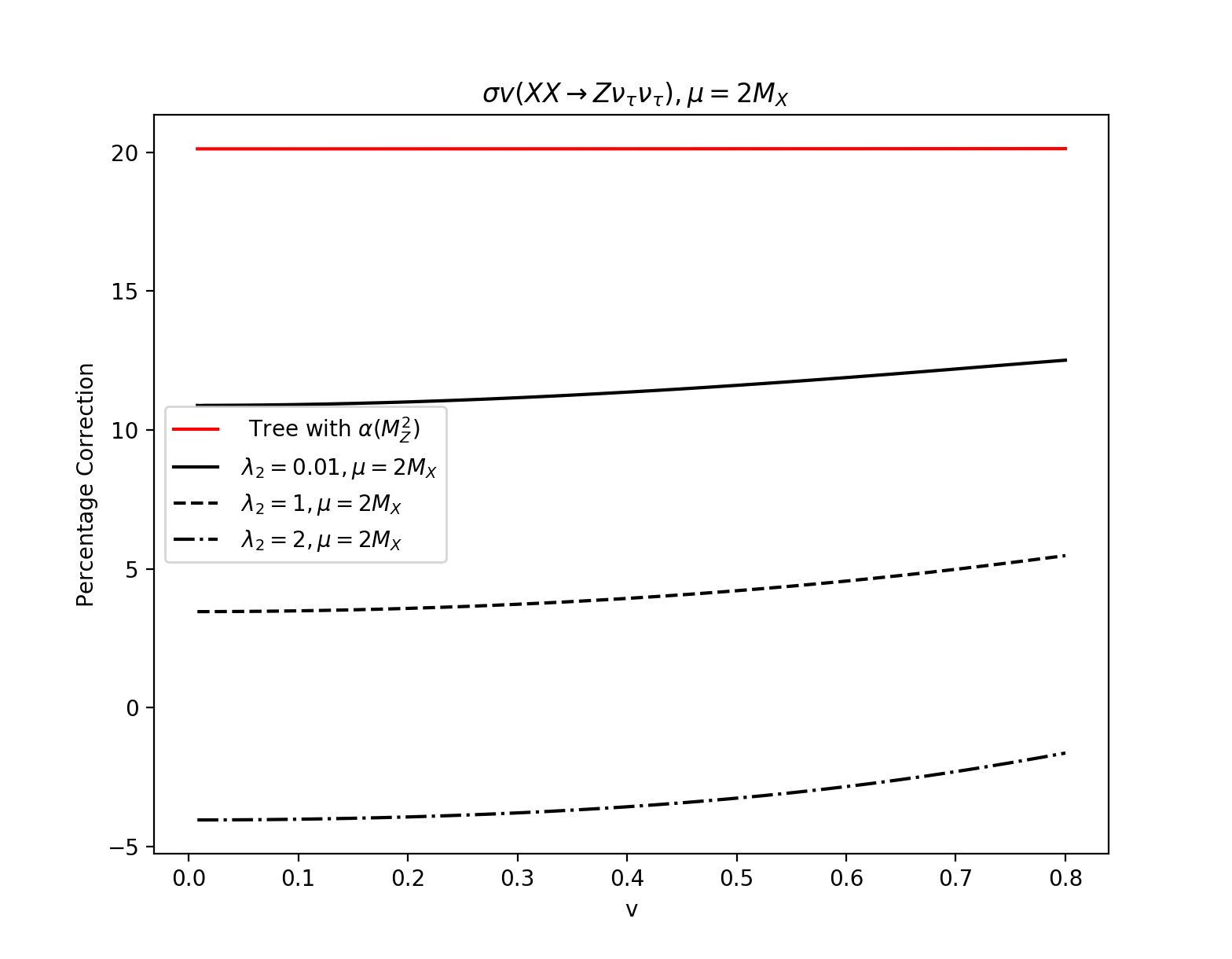}
\caption{\label{fig:relativecorrectionXXZnunubptI}\it Point G. Comparing the relative correction in $\%$ as a function of the relative velocity for $Z \nu_\tau \bar \nu_\tau$ for $\l_2=0.01,1,2$ and for 3 scales $\mu=M_X/,M_X,2M_X$. We also display the {\it improved tree-level} based on the use of $\alpha(M_Z^2)$. The panel on the right is a zoom on the choice $\mu=2 M_X$ for better readability.
}
\end{center}
\end{figure}
\end{center}

The results of the full one-loop corrections for three values of $\l_2=0.01,1,2$ and for different scales, $\mu=M_X/2,M_X,2 M_X$ are displayed in Figure~\ref{fig:relativecorrectionXXZnunubptI} for the range of relative velocities of interest for the relic density calculation. The so-called improved tree-level based on the use of $\alpha(M_Z^2)$ gives a constant correction of about $21\%$. $\mu=M_X/2$ gives not only the largest correction but shows also a significant velocity dependence. $\mu=M_X/2$ is not an appropriate scale, this scale is quite removed from the (largest) scales that enter the loop integrals: the invariant mass of the $XX$ system $\sqrt{s_{XX}} \simeq 2 M_X (1+v^2/8) \sim  2M_X$ or $M_A$ that enter in the $t$-channel exchange. As discussed in~\cite{OurPaper1_2020}, the appropriate scale should be, $\text{max}(2M_X,M_A)=158$ GeV. For point G, there is a small difference of 14 GeV between $2 M_X$ and $M_A$. We will come back to the choice $\mu=M_A$ whose results will be close to what we obtain for $2M_X$. Our results show that for $\mu=2 M_X$ and $\l_2=0.01$, the correction is about $11\%$. It decreases slowly as $\l_2$ increases. With this choice of the scale, the corrections range from $11\%$ to $-3\%$ for $\l_2$ ranging from $0.01$ to $2$. Observe that while for $\l_2=0.01$ the corrections are closest to the value obtained with the {\it improved tree-level} cross-section ($\alpha(M_Z^2)$), there is still as much as $10\%$ difference between the two corrections. An important lesson is that the $\l_2$ dependence of the full one-loop correction is clearly important. This $\l_2$ dependence is not all contained in $\bll$. 

\subsection{$X X\to Z \nu \bar{\nu}$ at Points F and A} 
Benchmark points A and F show similar trends to what we just saw for Point G despite the fact that both benchmarks points A and F have a much larger value of $\l_L$. This is understandable since the crucial property that explains the scale dependence is on the one-hand the relative $\l_L$ dependence on the cross-section and on the other the value of $\bll$ and its $\l_2$ dependence. The $\l_L$ dependence of the tree-level cross-section is not very different from that of point G. What is quite different is the magnitude of $\beta_L$ and its $\l_2$ dependence. Point F (A) has a smaller (larger) $\beta_L$ (about a factor 2 for the same value of $\l_2$) than Point G. When selecting the most appropriate scale we observe that for point A, $M_A$ is the most appropriate scale while for point F, $2 M_X$ is quite close to $M_A$. 
\begin{center}
\begin{figure}[tbhp]
\begin{center}
\includegraphics[width=0.65\textwidth, height=0.4\textwidth]{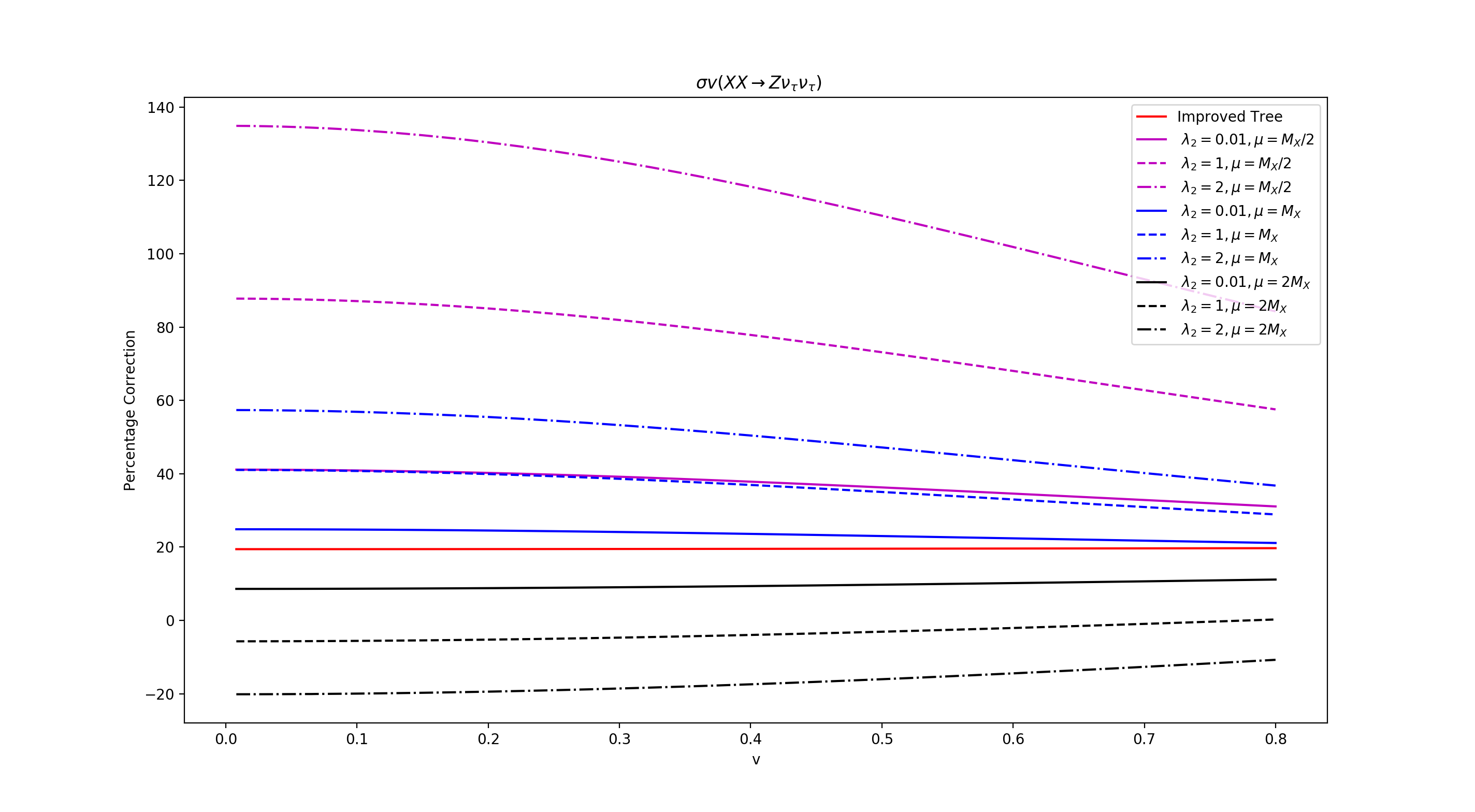} \hspace*{-0.5cm}
\includegraphics[width=0.34\textwidth, height=0.4\textwidth]{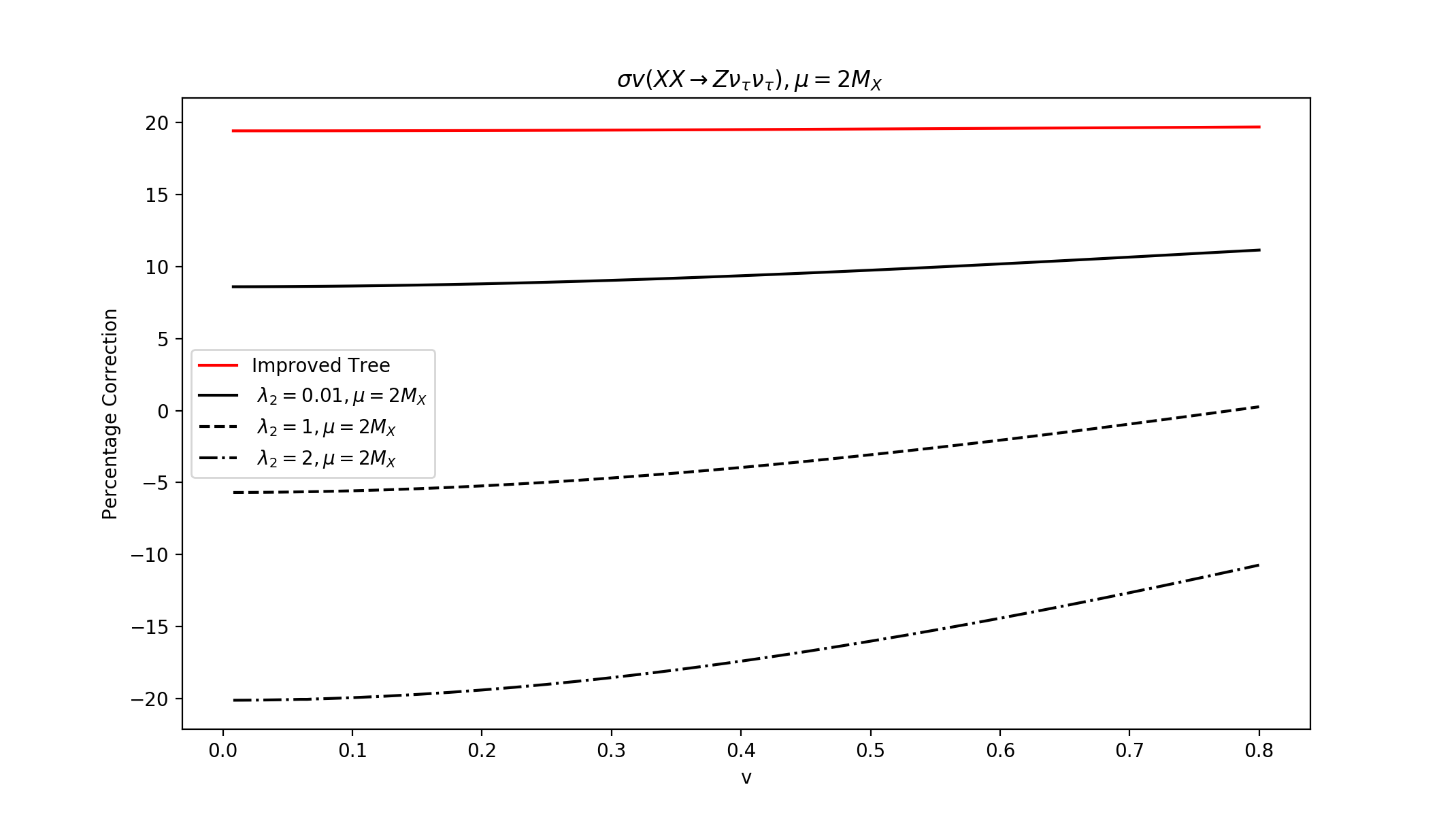}
\includegraphics[width=0.65\textwidth, height=0.4\textwidth]{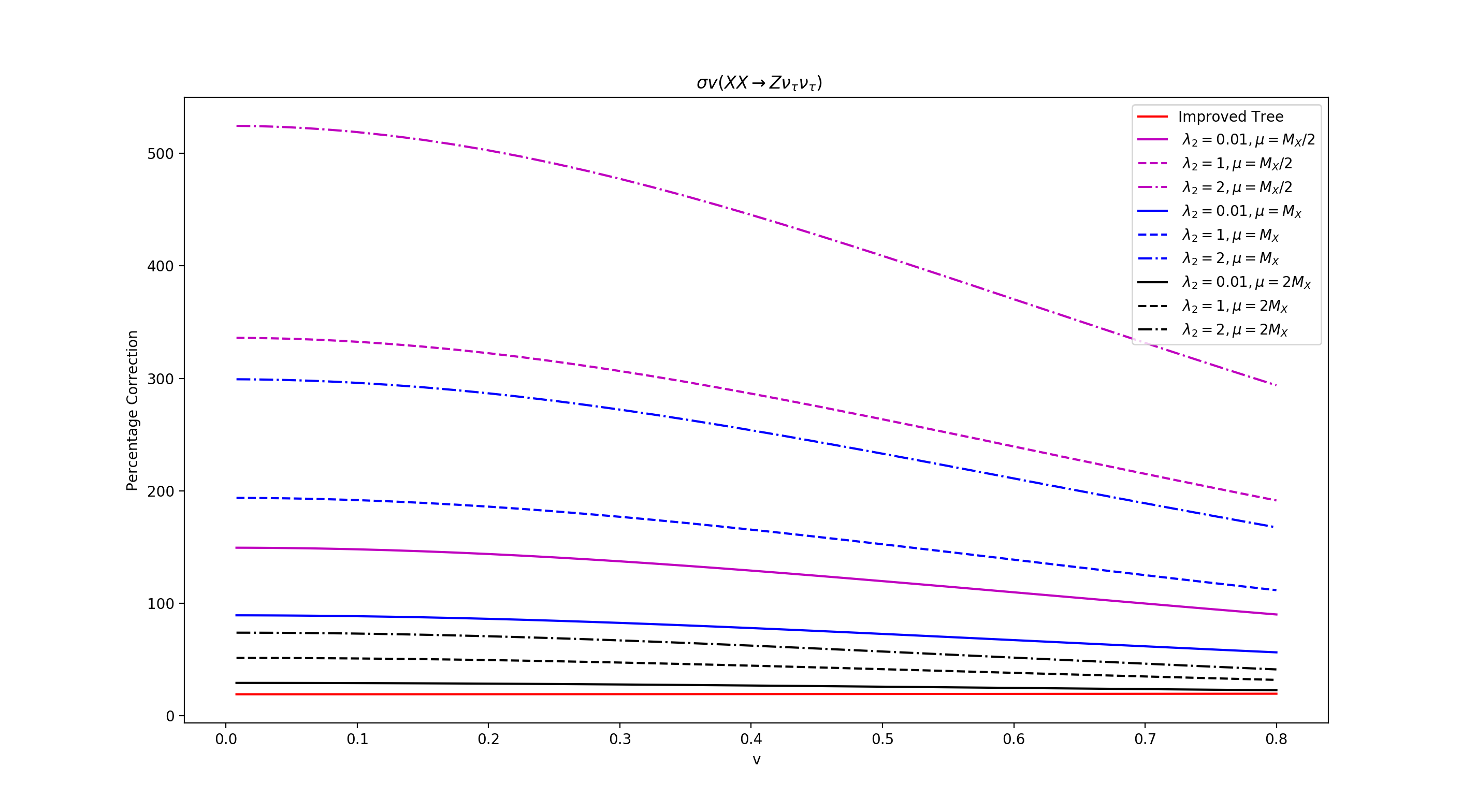} \hspace*{-0.5cm}
\includegraphics[width=0.34\textwidth, height=0.4\textwidth]{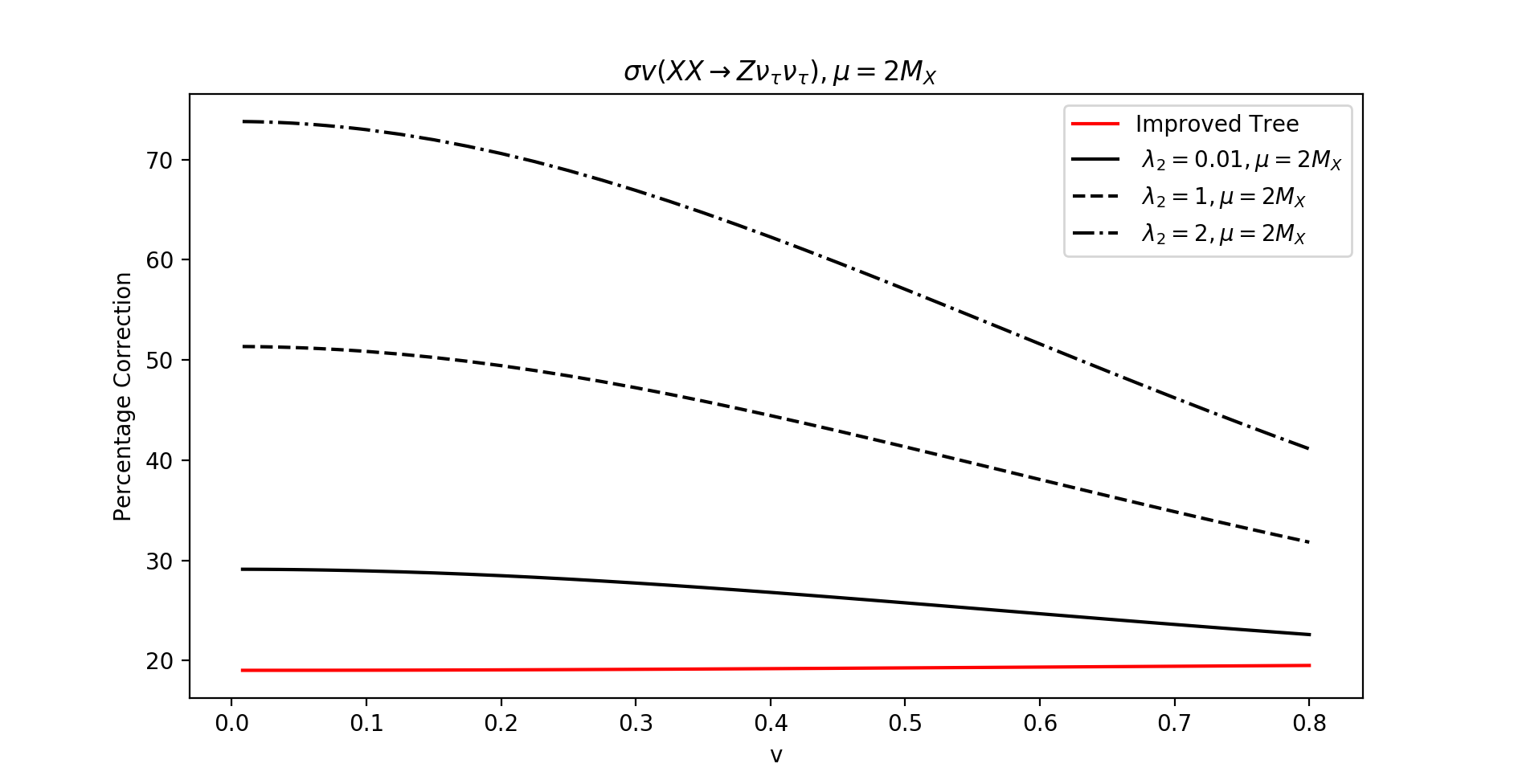}
\caption{\label{fig:relativecorrectionXXZnunubptHB}\it As in~\ref{fig:relativecorrectionXXZnunubptI} but for Point F (upper panels) and Point A (lower panels).}
\end{center}
\end{figure}
\end{center}
From the $\l_L$ dependence of the tree-level cross-section of point F, at $v=0.4$, we have 
\beqn
\sigma_{XX\to Z \nu \bar \nu, F}^0(v=0.4)\; v \simeq  0.012 \; (1+24 \l_L+143\l_L^2),
\eeqn 
so that the percentage correction is 
\beqn
\label{eq:sigmazzstarmudepF}
\left. \frac{{\rm d}\sigma_{XX\to Z \nu \bar \nu}( \mu_2)}{\sigma^0_{XX\to Z \nu \bar \nu}} \right\rvert_{F,v=0.4}\simeq \Bigg(\left. \frac{{\rm d}\sigma_{XX\to Z \nu \bar \nu}(\mu_1)}{\sigma^0_{XX\to Z \nu \bar \nu}} \right\rvert_{F, v=0.4}\;-\;20.1 \;(1+1.9\l_2)\; \ln(\mu_2/\mu_1)\Bigg)\%
\eeqn
while for point A, we have 
\beqn
\sigma_{XX\to Z \nu \bar \nu, A}^0(v=0.4)\;v \simeq 0.009 \; (1+28 \l_L+201\l_L^2),
\eeqn 
giving
\beqn
\label{eq:sigmazzstarmudepB}
\left. \frac{{\rm d}\sigma_{XX\to Z \nu \bar \nu}(\mu_2)}{\sigma^0_{XX\to Z \nu \bar \nu}} \right\rvert_{A,v=0.4} \simeq \Bigg(\left. \frac{{\rm d}\sigma_{XX\to Z \nu \bar \nu}(\mu_1)}{\sigma^0_{XX\to Z \nu \bar \nu}} \right\rvert_{A, v=0.4}\;-\;72.1 \; (1+1.4\l_2) \; \ln(\mu_2/\mu_1)\Bigg)\%
\eeqn

\begin{center}
\begin{figure}[tbhp]
\begin{center}
\includegraphics[width=0.34\textwidth, height=0.4\textwidth]{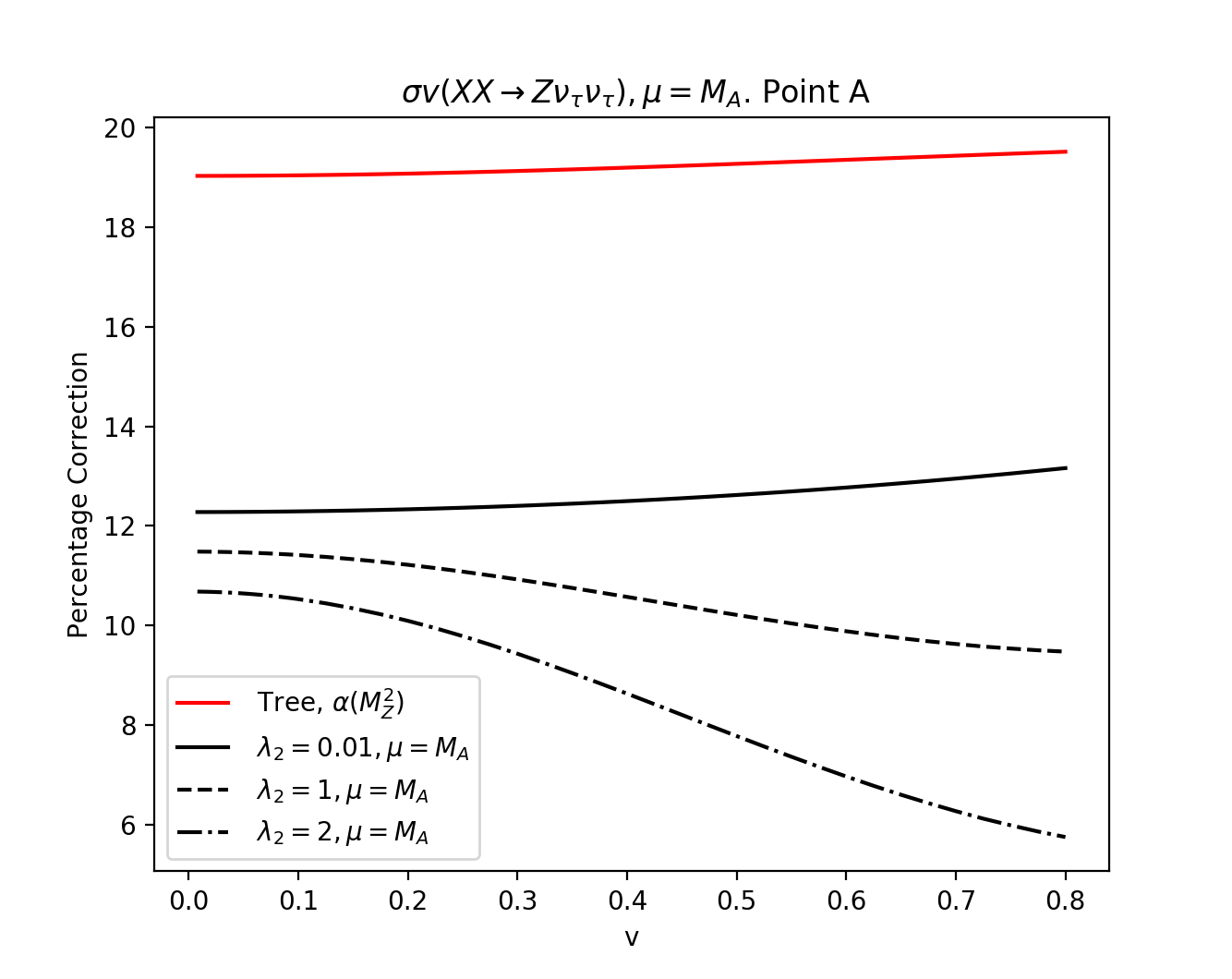}
\includegraphics[width=0.34\textwidth, height=0.4\textwidth]{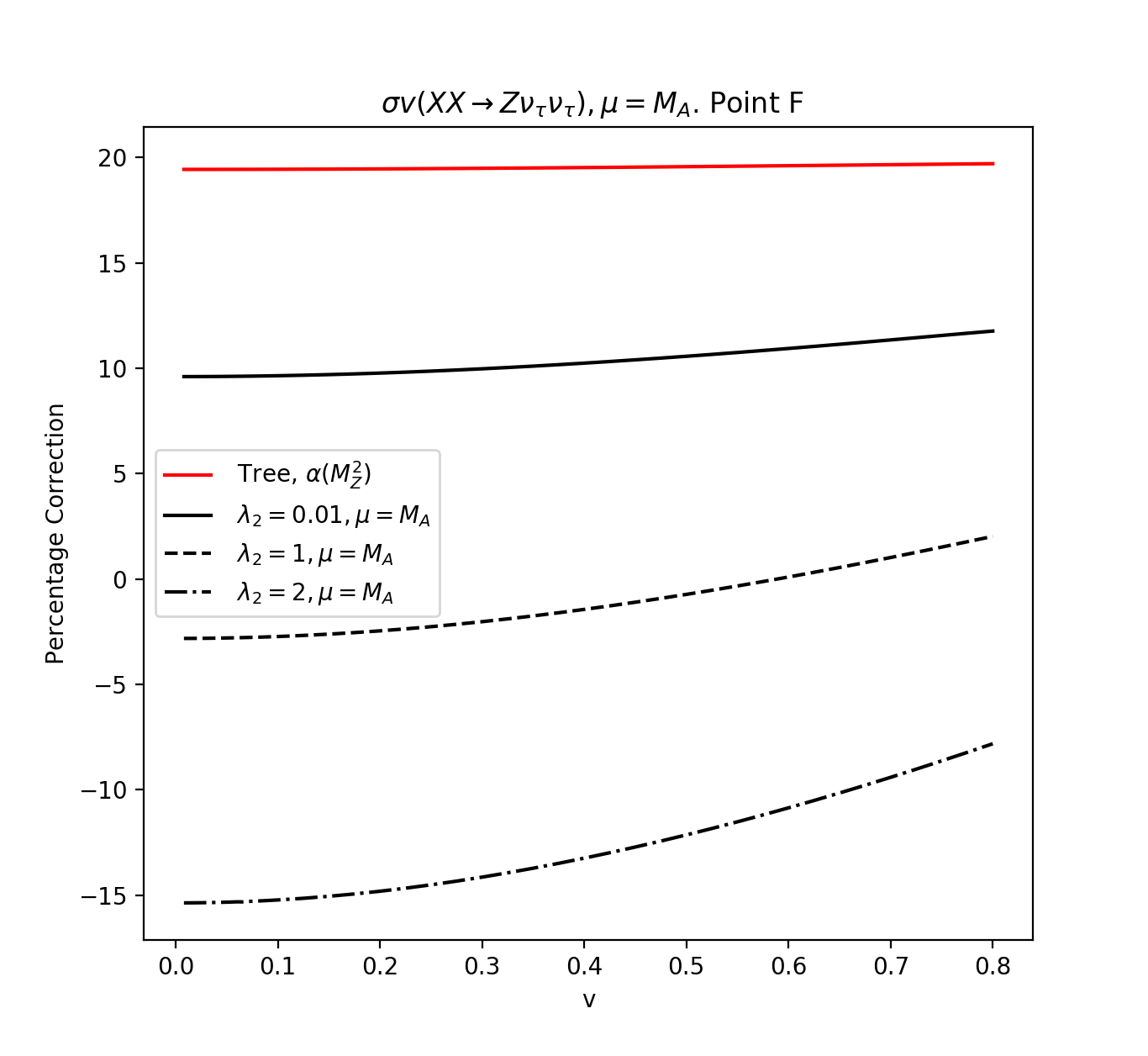}
\caption{\label{fig:relativecorrectionXXZnunubptHBMA}\it The relative corrections for $\sigma v(XX \to Z\nu_\tau \bar \nu_\tau)$ with $\mu=M_A$ for the benchmark Point A (left panel) and Point F (right panel).}
\end{center}
\end{figure}
\end{center}

Our results for points F and A are shown in Figure~\ref{fig:relativecorrectionXXZnunubptHB}. They confirm the general trend observed for point G. The full one-loop results displaying the scale dependence and $\l_2$ dependence are also in excellent agreement with the analytical formulae of Equations~\ref{eq:sigmazzstarmudepF}-\ref{eq:sigmazzstarmudepB}. For $\mu=M_X/2$, the corrections are very large with strong velocity dependence. Of the three scales, $\mu=M_X/2,M_X,2 M_X$, the largest scale, $\mu=2 M_X$, is where the corrections are the smallest.  Yet, for point A, even $\mu=2M_X$ gives large correction. For point A, $M_A$ is the largest scale for the process. It is quite different from the choice $\mu=2M_X$, considering the large value of $\bll$. We therefore show the results of taking $\mu=M_A$ as the optimised scale in Figure~\ref{fig:relativecorrectionXXZnunubptHBMA}. For Point F, the difference with $\mu=2M_X$ is very small and changes (slightly) mainly the results for $\l_2=2$ (there is an upward  shift of $-5\%$ when moving from $\mu=2M_X$ to $\mu=M_A$). This is due to the fact that for point F, $2M_X$ and $M_A$ are very close ($M_A=138$ GeV and $2M_X=144$ GeV) and $\bll$ is not large. The effect of the change of $\mu$ for point A when moving between the scale $2M_X$ and $M_A$ is substantial. In particular, for $\l_2=0.01$, the correction of $10\%$ is in line with the correction found for the 2 other benchmarks points while the corrections for $\l_2=1,2$ are sensibly smaller. Nonetheless, we warn the reader again that a large value of $\beta_{\l_L}$ is sensitive to large scale variations. Independently of the scale choice, a common feature is that $\l_2=0.01$ gives the smallest correction often approaching the tree-level improved $\alpha(M_Z^2)$ approximation. 


\section{$X X\to Z f \bar f$ and $X X \to W f \bar f^\prime$: one-loop results}
\label{sec:XXtoVffone-loop}
We already saw in our study of the tree-level cross-sections in section~\ref{sec:treelevel2to3} that the flavour dependence of $\sigma(X X \to W f \bar f^\prime)$ is, to an excellent accuracy, contained in and represented by the flavour dependence of $\Gamma(W \to f \bar f^\prime)$. We also learnt that the velocity dependence of the neutral and charged channels cancels out in the ratio of the cross-sections $\sigma(X X \to W f \bar f^\prime)/\sigma(X X\to Z \nu  \bar \nu)$ for velocities up to $v=0.5$. Above these velocities, the $WW$ channel starts experiencing the onset of the $WW$ threshold while the $ZZ$ is still not experiencing the $ZZ$ threshold. The underlying global $SU(2)$ symmetry would also suggest that, particularly below $v<0.5$, the $\l_L$ dependence of $\sigma(X X \to W f \bar f^\prime)/\sigma(X X\to Z \nu  \bar \nu)$ cancels out. To wit, we find that the $\l_L$ dependence for $XX \to W \nu_\tau \bar \tau$ for point G and velocity $v=0.4$ writes as 
\beqn
 \sigma_{XX\to W \nu_\tau \bar \nu_\tau, G}^0(v=0.4)&=&0.086 + 1.960 \l_L + 11.343 \l_L^2 \simeq 0.086\; (1+ 23 \l_L + 133 \l_L^2) \nonumber \\
 & \simeq & 6.585 \; \sigma_{XX\to Z \nu \bar \nu, G}^0(v=0.4). 
 \eeqn
The latter very good approximation means, especially for the very small values of $\l_L$ we are permitted, that the $\mu$ dependence, Equation~\ref{eq:sigmazzstarmudepI}, is within machine precision and fitting procedure, identical for $Z Z^\star$ and $W W^\star$. Since these two channels carry almost the same relative $\l_L$ dependence and that $\l_L\ll 1$, the scale dependence for the normalised cross-sections is confirmed to be almost the same. A small expected departure above $v> 0.5$ is confirmed numerically. The flavour independence of $\sigma(X X \to W f \bar f^\prime)/\Gamma(W \to f \bar f^\prime)$ continues to hold true at one-loop. The latter stems from the fact that the electroweak radiative corrections (relative to the tree-level) of $\Gamma(W \to f \bar f^\prime)$ are known to be the same for all flavours~\cite{Bardin:1986fi}, therefore $R^{XX}_{W f \bar f^\prime} =1$ at one-loop also. We also verify all these properties by a direct full one-loop computation to the different channels. First of all, we confirm that the relative electroweak correction to the annihilation into $W l \bar{\nu}_l$ is, within the per-mille level, the same as that of the annihilation into $W q \bar{q}^\prime$, we therefore only show the leptonic (charged) final state, $W \tau \bar{\nu}_\tau$.
\begin{center}
\begin{figure}[htbp]
\begin{center}
\includegraphics[width=0.64\textwidth, height=0.45\textwidth]{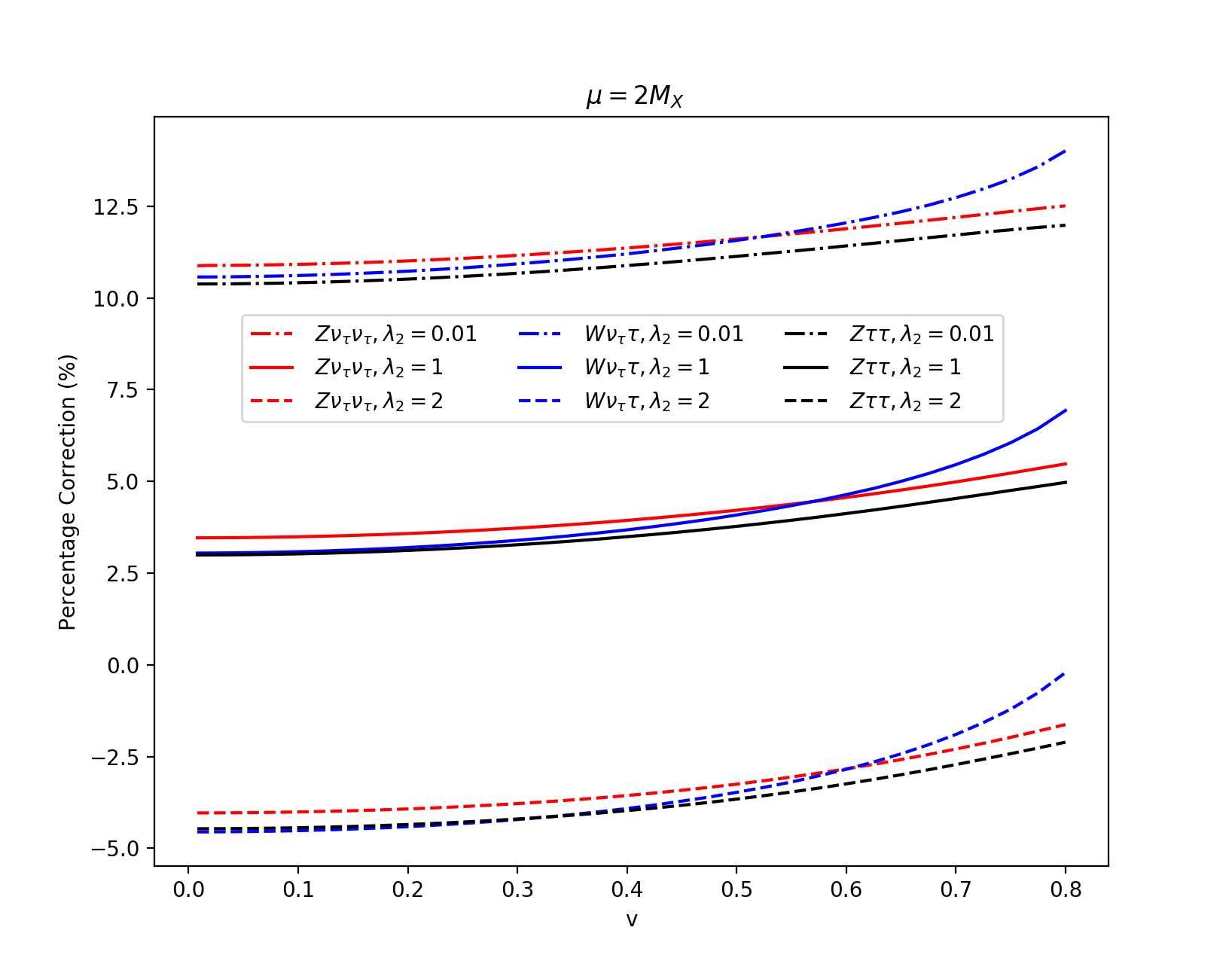}
\includegraphics[width=0.34\textwidth, height=0.45\textwidth]{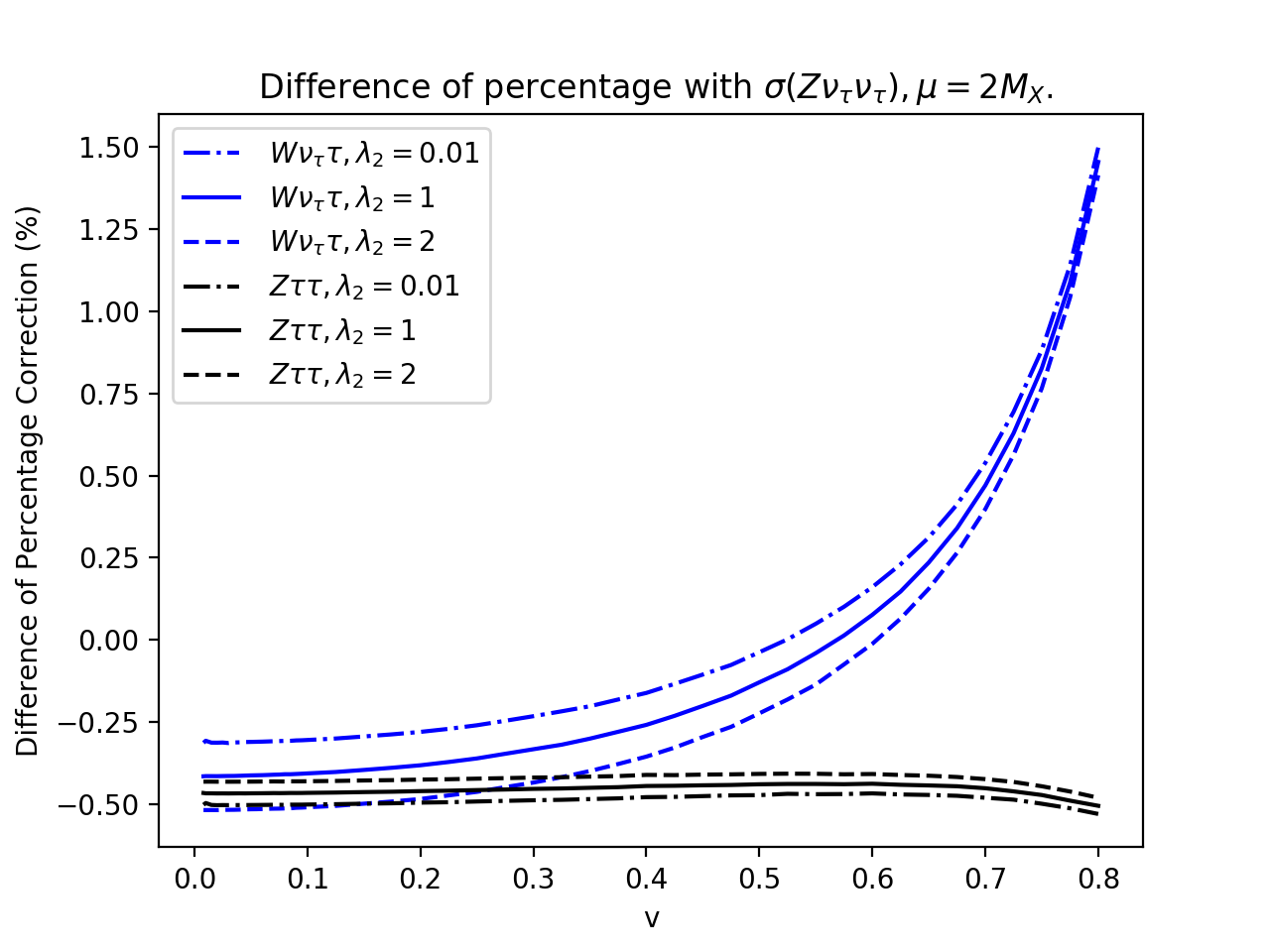}
\caption{\label{fig:relativecorrectionmu2mxI}\it Point G. The relative correction, ${\rm d}\sigma(V f \bar{f}^\prime)/\sigma^0(V f \bar{f}^\prime), V=W, Z$ in $\%$ as a function of the relative velocity for $XX \to Z \tau^+ \tau^-, Z \nu_\tau \bar \nu_\tau, W \tau \bar \nu_\tau$ for $\mu=2 M_X$ and  $\l_2=0.01,1,2$. The right panel shows the difference ${\rm d}\sigma(V f \bar{f}^\prime)/\sigma(V f \bar{f}^\prime)-{\rm d}\sigma(Z \nu \bar \nu)/\sigma(Z \nu \bar \nu)$.}
\end{center}
\end{figure}
\end{center}
As Figure~\ref{fig:relativecorrectionmu2mxI} for $\mu=2M_X$ in the benchmark point G shows, the behaviour of all the annihilation channels follow $XX \to Z \nu_\tau \bar \nu_\tau$. The $\l_2$ dependence is indeed represented by the channel we discussed in detail in the previous section. The $-0.5\%$ difference between the $(Z) \tau \bar \tau$ and $(Z) \nu_\tau \bar{\nu}_\tau$ is indeed due mostly to the one-loop relative correction between the $Z$ decay width into $\tau \bar \tau$ and $\nu_\tau \bar{\nu}_\tau$ ($-0.9\%$) with a smaller contribution from the $\tau$ Yukawa mass ($h$ exchange for example) as identified in Figure~\ref{fig:relativeweightXXVVstar}. The deviation observed in $ W\tau \bar{\nu}_\tau$ increases for larger $v$ (as expected from the tree-level comparison of these two channels). Nonetheless, the difference between the \underline{relative} corrections remains below $1.5\%$ (below $0.5\%$ for $v<0.5$). To an excellent approximation, the relative one-loop corrections, ${\rm d}\sigma$, between the different channels, represent the difference between the one-loop electroweak corrections to the corresponding partial widths, ${\rm d} \Gamma$,

 \beqn
 \label{eq:diffsiggamf}
 \frac{{\rm d} \sigma(XX \to Z f \bar f) v}{ \sigma(XX \to Z f \bar f) v} -  \frac{{\rm d} \sigma(XX \to Z \nu \bar \nu ) v}{ \sigma(XX \to Z \nu \bar \nu) v} \simeq 
 \frac{{\rm d} \Gamma(Z \to f \bar f)}{ \Gamma(Z \to f \bar f)} - \frac{{\rm d} \Gamma(Z \to \nu \bar \nu)}{ \Gamma(Z \to \nu  \bar \nu)}  
 \eeqn

\begin{center}
\begin{figure}[bthp]
\begin{center}
\includegraphics[width=0.32\textwidth, height=0.4\textwidth]{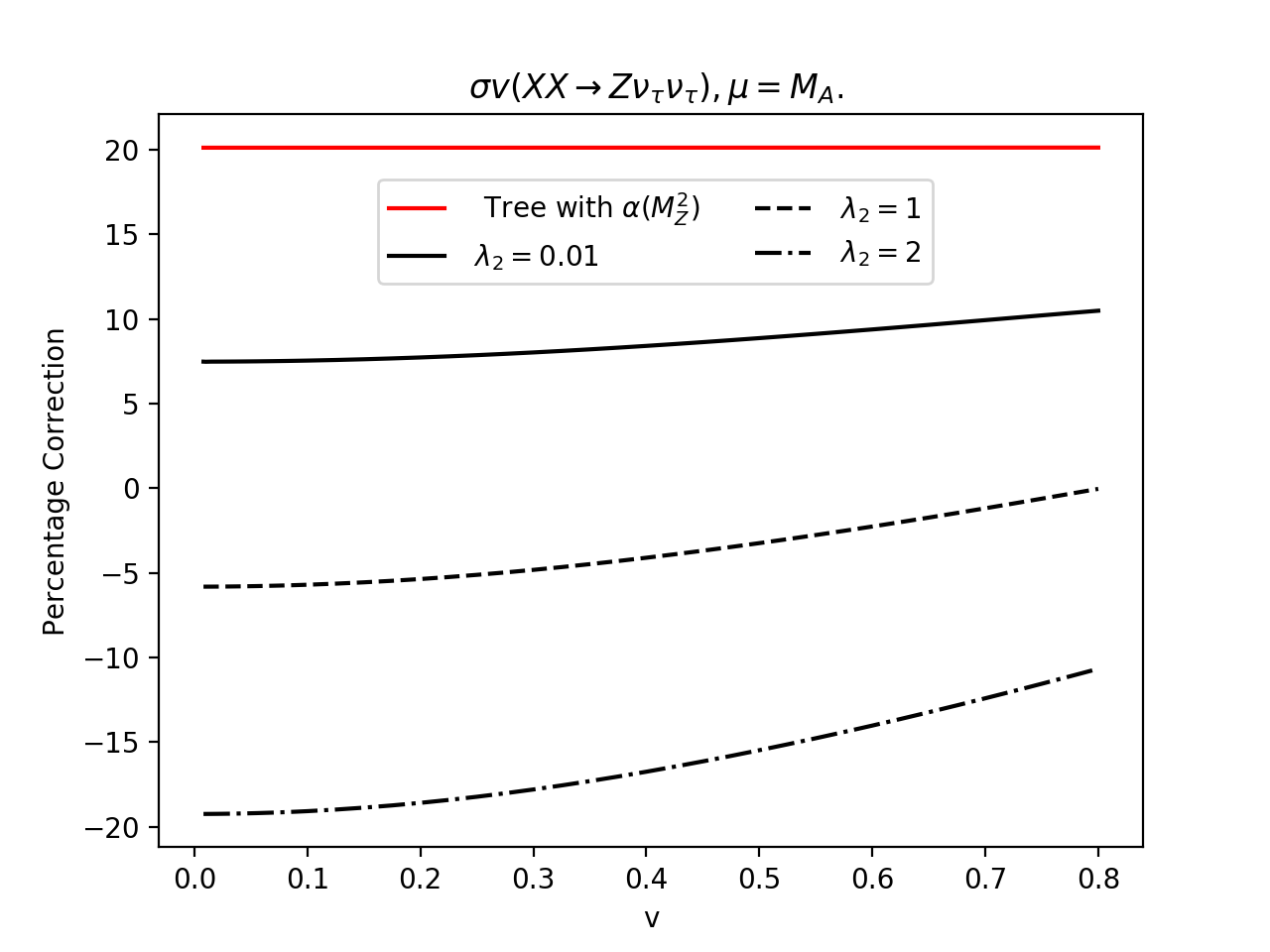}
\includegraphics[width=0.32\textwidth, height=0.4\textwidth]{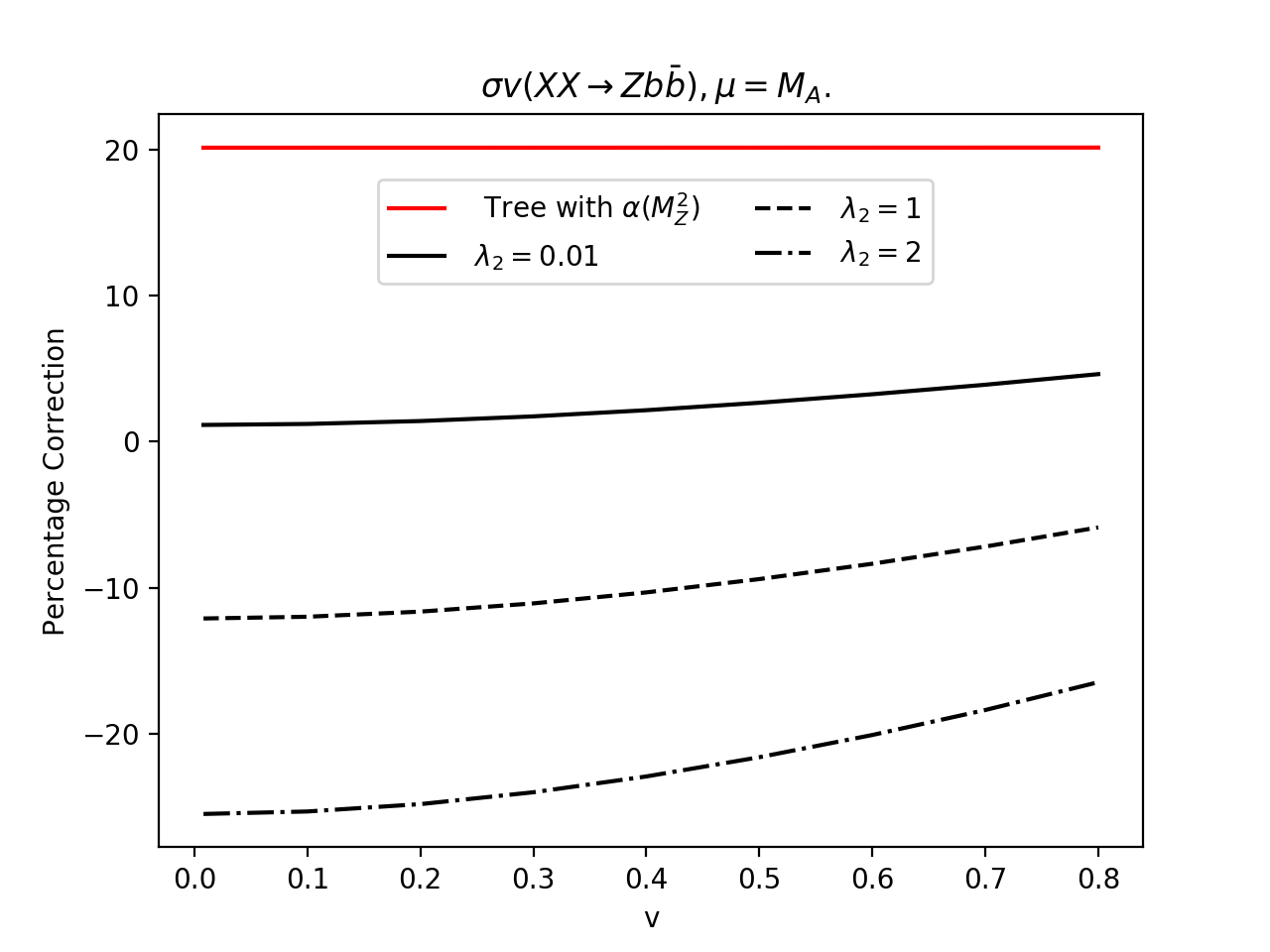}
\includegraphics[width=0.32\textwidth, height=0.4\textwidth]{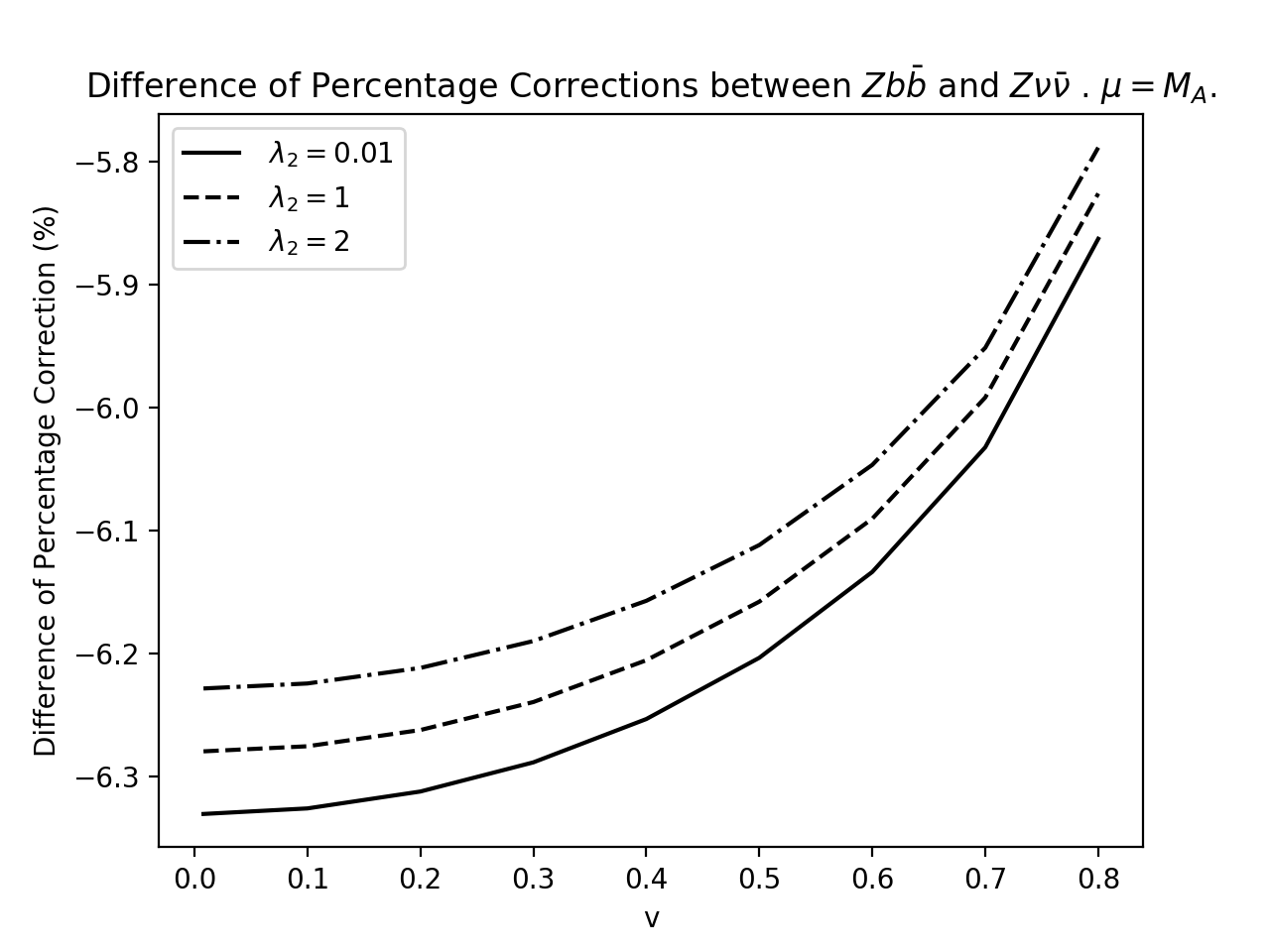}
\caption{\label{fig:relativecorrectionXXZnunubbarptI}\it Point G. Percentage corrections for $\mu=M_A$ for the $Z \nu_\tau \bar \nu_\tau$ and $Z b \bar b$ final states. The latter are about $6\%$ lower than the neutrino case for all values of $v$ and $\l_2$.
}
\end{center}
\end{figure}
\end{center}

\noi The mass effect is obviously more important for the $b$-quark final state channel, as we already saw at tree-level (Figure~\ref{fig:relativeweightXXVVstar}). In the radiative corrections to $\sigma v (XX \to Z b \bar b)$, there is about a $-6\%$ difference with the $Z \nu \bar \nu$ channel, see Figure~\ref{fig:relativecorrectionXXZnunubbarptI}. Almost half of this correction is due to the difference in the relative electroweak correction between the partial width $Z \to b \bar b$ and $Z \to \nu \bar \nu$. 


\section{Intermediate Summary}
\label{sec:intermediatesummary}
Let us take stock. 
\begin{itemize}
\item An important feature common to all three benchmarks is that whatever the values of the renormalisation scale $\mu$ and of the parameter $\l_2$ are, $XX \to Z \nu \bar \nu$ encapsulates practically all of the radiative corrections contained not only in the neutral channels $XX \to Z f \bar f$ but also in the charged channels $XX \to W f \bar f^\prime$, in the sense that the normalised one-loop corrections are, to a high degree of accuracy, equal for all channels for the same choice of $\l_2$ and the renormalisation scale $\mu$ 
\beqn
\label{eq:summ_ww_zz}
\frac{\delta \sigma(XX \to Z \nu \bar \nu)}{\sigma(XX \to Z \nu \bar \nu)}{\Huge |}_{\mu, \l_2} \sim \frac{\delta \sigma(XX \to V f \bar f^\prime)}{\sigma(XX \to V f \bar f^\prime)}{\Huge |}_{\mu, \l_2} \quad V=W,Z
\eeqn
\subitem - Equation~\ref{eq:summ_ww_zz} is valid at all $v$ (in the range of interest for the relic density) for $\sigma(XX \to Z f \bar f)$. There is a slight $v$ dependence for $\sigma(XX \to W f \bar f^\prime)$ when compared to $\sigma(XX \to V f \bar f)$. This difference is small and does not exceed more than $1.5\%$.
\subitem -There is no flavour dependence for the charged channels. The flavour dependence in the neutral channel is largest for $Zb\bar b$ where the largest difference amounts to $6\%$, more than half of this difference is accounted for by a correction given by $\Gamma(Z \to f \bar f)_{{\rm 1-loop EW}}/\Gamma(Z \to f \bar f)_{{\rm tree}}$. We expect these small flavour effects to be diluted when we consider the correction to the relic density, considering that the $ZZ$ channel accounts for about $10\%$ to the relic density and that the $b \bar b$ channel is $15\%$ of the whole $ZZ$. 
\item As expected, the scale dependence is largest for point A which has the largest $\bll$. Our conjecture (based nonetheless on the study of the scales involved in the loop functions) seems to be a very good one. The appropriate scale is ${\rm max}(2M_X,M_A)$. In particular, we find that for $\l_2=0.01$, all 3 benchmark points give very similar corrections of about $10\%-12\%$. For this choice of scale, points F and G (which have small $\bll$) give very similar corrections for $\l_2=1 \; (-5\%) ,2 \; (-20\%)$ while for point A, the corresponding corrections are $v$ dependent with values for $v\sim 0$ similar for all $\l_2$. We observe that in our study of $XX \to W f \bar f^\prime$ (P60) in the co-annihilation region~\cite{OurPaper2_2020}, where an on-shell renormalisation for $\l_L$ was possible, the electroweak correction for $\l_2=0.01,1,2$ were {\it quantitatively} very similar to the results we obtain here, especially for the benchmark points F and G. This validates further our conjecture about the choice of scale. For the calculation of the relic density, we consider that the appropriate scale is  ${\rm max}(\mu=2M_X,M_A)$ and that theoretical uncertainty can be estimated by the difference within the range $(2M_X,M_A)$. 
\end{itemize}


\section{Effect on the relic density}
\label{sec:reliconeloop}
\begin{table}[htb]
\begin{center}
\begin{tabular}{|c||c|c|c|c|c|}
\cline{2-6}
\multicolumn{1}{c|}{}& tree&$\alpha(M_Z^2)$ & $\l_2=0.01$ & $\l_2=1$ & $\l_2=2$\\
\hline \hline
{\bf Point G (I)} & 0.121 &0.101 & & & \\
& & {\tiny (-16.53\%)}& & & \\
Full, $\mu=M_X$ & & & 0.093 {\tiny (-23.43\%)} & 0.076 {\tiny (-36.73\%)} & 0.065 {\tiny (-46.11\%)} \\
Full, $\mu=2 M_X$ & & & 0.109 {\tiny (-9.53\%)} & 0.117 {\tiny (-3.47\%)} & 0.125 {\tiny (3.57\%)} \\
Full, $\mu=M_A$ & & & 0.112 {\tiny (-7.27\%)} & 0.126 {\tiny (3.94\%)} & 0.143 {\tiny (18.51\%)} \\ \hline
Simplified, $\mu=M_X$ & & & 0.093 {\tiny (-23.48\%)} & 0.076 {\tiny (-36.76\%)} & 0.065 {\tiny (-46.12\%)} \\
Simplified, $\mu=2 M_X$ & & & 0.109 {\tiny (-9.60\%)} & 0.117 {\tiny (-3.55\%)} & 0.125 {\tiny (3.47\%)} \\
Simplified, $\mu=M_A$ & & & 0.112 {\tiny (-7.35\%)} & 0.126 {\tiny (3.84\%)} & 0.143 {\tiny (18.37\%)} \\ \hline \hline
{\bf Point A (B) } & 0.156 & 0.130 & & &\\
& & {\tiny (-16.67\%)} & & &\\
Simplified, $\mu=M_X$ & & & 0.092 {\tiny (-41.35\%)} & 0.063 {\tiny (-59.57\%)} & 0.048 {\tiny (-69.12\%)} \\
Simplified, $\mu=2 M_X$ & & & 0.125 {\tiny (-19.59\%)} & 0.112 {\tiny (-28.28\%)} & 0.101 {\tiny (-35.31\%)} \\
Simplified, $\mu=M_A$ & & & 0.140 {\tiny (-10.16\%)} & 0.144 {\tiny (-7.91\%)} & 0.147 {\tiny (-5.52\%)} \\ \hline \hline
{\bf Point F (H)} & 0.119 &0.099 & & & \\
& & {\tiny (-16.81\%)} & & & \\
Simplified, $\mu=M_X$ & & & 0.098 {\tiny (-17.92\%)}& 0.089 {\tiny (-25.23\%)} & 0.082 {\tiny (-31.38 \%)} \\
Simplified, $\mu=2 M_X$ & & & 0.109 {\tiny (-8.15\%)}& 0.123 {\tiny (3.62\%)} & 0.142 {\tiny (19.17\%)} \\
Simplified, $\mu=M_A$ & & & 0.108 {\tiny (-8.82\%)}& 0.120 {\tiny (1.21\%)} & 0.136 {\tiny (13.94\%)} \\ \hline \hline
\end{tabular}
\end{center}
\caption{\it The relic density for points G, A and F at tree-level and after including the one-loop corrections. The percentage changes are given in parenthesis. The percentage correction corresponding to the use of $\alpha(M_Z^2)$ at tree-level is also indicated. The full corrections for point G differ from the simplified one-loop in that the full one-loop $Zb\bar b$ final state is fully taken into account while in the simplified version all fermion final states are rescaled from the full one-loop in the $Z \nu \bar \nu$ cross-section through the added one-loop flavour correction, $\Gamma_{Z\to f \bar f}/\Gamma_{Z\to \nu \bar \nu}$, where the partial widths are computed at one-loop, see text for details.}
\label{tab:rel-cont-A_F_G}
\end{table}
We just learnt that the $v$ dependence of the cross-sections, that contributes to the relic density calculation, is rather smooth. Moreover, the $\mu$ (scale) dependence is sensibly the same in all channels (neutral and charged). We therefore expect the $\mu$ dependence of the relic density to follow that of the cross-section $\sigma(XX \to Z \nu \bar \nu)$, since $\Omega h^2 \sim 1/<\sigma v>$ ($<\sigma v>$ is the total thermodynamically averaged cross-section). The difference between the values of the relic density for $\mu=M_X$ and $\mu=2M_X$ follows this trend as shown in Table~\ref{tab:rel-cont-A_F_G} of the relic densities obtained after passing all our tree and one-loop $v$ dependent cross-sections to {\tt micrOMEGAs}. The table shows (as expected) large corrections for the inappropriate choice $\mu=M_X$, particularly for point A. We derive the relic density by taking into account the Yukawa couplings of the $b$-quarks (full calculation) beyond the effect of the flavour dependence contained in the partial decay $\Gamma_{Z \to b \bar b}$, see~\ref{eq:diffsiggamf}, that allows a nice factorisation of the total cross-section in terms of $\sigma (XX \to Z \nu \bar \nu)$, which we call {\it simplified}. The difference between the full and simplified implementations is very small since the overall contribution of the $Z b \bar b$ final state to the total annihilation cross-section is small. An important feature seen for all scales and benchmark points is that the impact of $\l_2$ is large, this parameter is not taken into account when tree-level analyses are conducted.\\

The appropriate scale is ${\rm max}(\mu=2M_X,M_A)$. For $\l_2=0.01$, the 3 benchmark points give very similar results with small corrections contained in the range $-7\%$ to $-10\%$. These corrections are smaller than those found through the naive use of a running $\alpha$ at scale $M_Z^2$. Even for these scales, the $\l_2$ dependence is not negligible at all. To give a quantitative estimate of the theory uncertainty that a tree-level evaluation of the relic density should incorporate, one needs to look at the one-loop results by varying both $\mu$ and $\l_2$. For instance, take the benchmark point F (very similar results are obtained for point G) which has a small $\bll$. While at tree-level, $\Omega h^2=0.119$ (obtained with $\alpha(0)$), the theory uncertainty now is $0.108<\Omega h^2<0.142$ ($\Omega h^2=0.119 ^{19.2\%}_{-8.8\%}$), this is more than the uncertainty of $\pm 10\%$ applied routinely in some analyses. Note that the uncertainty/error is much larger if based on the usage of $\alpha(M_Z^2)$ which is the default value of {\tt micrOMEGAs}. For point A, where $\bll$ is larger, the tree-level result is turned into the range $\Omega h^2=0.156^{-5.5\%}_{-35.3\%}$ with the conclusion that a value of $0.156$ that could be dismissed on the basis of the present experimental constraint on the relic density can in fact be easily brought in line with the measured value if loop corrections were taken into account. 


\section{Conclusions}
\label{sec:conclusions}
This is the first time a calculation of $2\to 3$ processes for the annihilation of DM has been performed at the one-loop level and the results of the corrected cross-sections turned into a prediction of the relic density. While this calculation is within the IDM, the tools at our disposal are now powerful enough to tackle such calculations for any model of DM provided a coherent renormalisation programme has been devised and implemented. In the particular case of the IDM, the reconstruction of the model parameters in order to fully define the model leaves two underlying parameters not fully determined in terms of physical parameters. One parameter, $\l_2$, describes, at tree-level, the interaction solely within the dark sector of the IDM. It is therefore difficult to extract {\it directly} from observables involving SM particles. Yet, this parameter contributes significantly to dark matter annihilation processes such as those we studied here. This indirect one-loop effect could in principle be extracted from the precise measurement of the relic density, a situation akin to the extraction of the top mass from LEP observables provided all other parameters of the model, masses of the additional scalars and their coupling to the SM, $\l_L$. $\l_L$, in fact measures the strength of the coupling of the SM Higgs to the pair of DM, there is in fact a one-to-one mapping between Higgs decay to $XX$ and $\l_L$, which suggests an extraction of $\l_L$ from the partial width of the Higgs into $XX$. While difficult in general, it is impossible when this Higgs decay is closed. The allowed parameter space for the $2\to 3$ processes we studied is when this Higgs decay is closed. In this case we suggested an $\overline{\text{MS}}$ scheme for $\l_L$. The $\overline{\text{MS}}$ introduces a scale dependence on the one-loop cross-sections. We showed that the scale dependence can be determined from the $\l_L$ parametric dependence of the tree-level cross-section and the knowledge of the 1-loop $\beta$ constant for $\l_L$, $\bll$. Despite the fact that the experimentally allowed values of $\l_L$ are small, the parametric dependence of the cross-sections on $\l_L$ are large for all the benchmarks that we studied. Combined with not so small $\bll$, the scale dependence can be very large if an {\it inappropriate} scale is chosen. Based on a few other analyses in the framework of the IDM~\cite{OurPaper2_2020} and also in supersymmetric scenarios~\cite{Belanger:2016tqb, Belanger:2017rgu} regarding the issue of the choice of the optimal scale, we suggest to restrict the choice of the scale to values around the maximal scale involved in the process. The present one-loop analysis is yet another warning to practitioners of the IDM and other BSM models in respect of the relic density of DM in these models. The one-loop analyses give an important quantitative estimate of the (often) large theoretical uncertainty that should be taken into account before allowing or dismissing scenarios based on a tree-level derivation of relic density. The latter for instance is not sensitive to the value of $\l_2$. $\l_2$ should be taken into account along side the uncertainty from the scale variation. We find that the combined theoretical uncertainty is model-dependent and in many cases is much larger than the cursory ($\pm$){\it symmetric} $10\%$ theoretical uncertainty that is included in many analyses.

\acknowledgments
We thank Alexander Pukhov for several insightful discussions. HS is supported by the National Natural Science Foundation of China (Grant , No.12075043, No.11675033). He thanks the CPTGA and LAPTh for support during his visit to France in 2019 when this work was started. SB is grateful for the support received from IPPP, Durham, UK, where most of this work was performed. SB is also grateful to the support received from LAPTh during his visit, when this work started. NC is financially supported by IISc (Indian Institute of Science, Bangalore, India) through the C.V.~Raman postdoctoral fellowship. NC also acknowledges the support received from DST, India, under grant number IFA19-PH237 (INSPIRE Faculty Award).

\bibliography{../../NLO_IDM570}
\bibliographystyle{JHEP}
\end{document}